\documentclass[8.5pt,twoside,twocolumn]{article}

\usepackage[super,sort&compress,comma]{natbib} 
\usepackage{mhchem}
\usepackage{times,mathptmx}
\usepackage{sectsty}
\usepackage{balance} 
\usepackage{graphicx} 
\usepackage{lastpage}
\usepackage[format=plain,justification=raggedright,singlelinecheck=false,font=small,labelfont=bf,labelsep=space]{caption} 
\usepackage{fancyhdr}
\usepackage{color}

\begin{document}





\twocolumn[
\begin{@twocolumnfalse}
\noindent\LARGE{\textbf{Amino Acids and Proteins at ZnO-water Interfaces in Molecular Dynamics Simulations}}
\vspace{0.6cm}

\noindent\large{\textbf{Grzegorz Nawrocki and Marek Cieplak}}\vspace{0.5cm}

{\small Institute of Physics, Polish Academy of Sciences,\\
Al. Lotnik\'ow 32/46, 02-668 Warsaw, Poland.\\
E-mail: mc@ifpan.edu.pl}

\vspace{0.5cm}



\noindent \normalsize{
We determine potentials of the mean force for interactions of amino acids 
with four common surfaces of ZnO in aqueous solutions. 
The method involves all-atom molecular dynamics simulations 
combined with the umbrella sampling technique. The profiled
nature of the density of water with the strongly adsorbed first
layer affects the approach of an amino acids to the surface 
and generates either repulsion or weak binding.
The largest binding energy is found for tyrosine interacting with 
the surface for which the Zn ions are at the top. It is equal to 7 kJ mol$^{-1}$
which is comparable to that of the hydrogen bonds in a protein. 
This makes the adsorption of amino acids to ZnO surface to be much weaker
than to the well studies surface of gold.
In vacuum, binding energies are more that 40 times stronger
(for one of the surfaces).
The precise manner in which 
water molecules interact with a given surface influences  the binding
energies in a way that depends on the surface.
Among the four considered surfaces the one with Zn at the top is recognized as
binding almost all amino acids with the average binding energy of
2.60 kJ mol$^{-1}$. Another (O at the top) is  non-binding for most amino acids.
For binding situations the average energy is 0.66 kJ mol$^{-1}$.
The remaining two surfaces bind nearly as many amino acids as they do not
and the average binding energies are 1.46 and 1.22 kJ mol$^{-1}$. 
For all of the surfaces the binding energies vary between amino acids 
significantly: the dispersion in the range of 68 - 154\% of the mean.
A small protein is shown to adsorb to ZnO only intermittently 
and with only a small deformation. Various adsorption events
lead to different patterns in mobilities of amino acids within the protein. 
}
\vspace{0.5cm}
\end{@twocolumnfalse}
]



\section{Introduction}

Understanding the nature of protein adsorption on solid surfaces
is a fundamental problem in biophysics \cite{hlady,gray,feng,nakanishi} 
which has important implications in biological materials and engineering.
They relate to many applications 
that include biosensing \cite{Lynch}, biomaterials \cite{hlady}, 
drug delivery \cite{haynes} and industrial chemistry \cite{somorjai}.
Some of these applications rely on recognition of the surfaces
by specific proteins or peptides. It is thus important to investigate
the role of the sequential makeup in binding to the surface.

Elucidation of interactions of a protein with a surface 
should start at the single  amino acid (AA) level.
Examples of questions that need to be answered are:
what is the characteristic energy of interactions, 
what is the level of specificity in the binding affinities, 
and what are the effects of the surrounding water molecules.
There are experimental studies of GaAs, InP, and Si suggesting 
that one should be able to design sequences 
of high binding specificity \cite{Whaley,Bachmann} to these surfaces. 

In particular, liquid chromatography studies of adsorption of AAs to silicon \cite{Basiuk} 
suggest that the AA-averaged lowering 
of the free energy due to adsorption is 1.15$\pm$1.05 kJ mol$^{-1}$ 
(it varies between -0.40 and 3.64 kJ mol$^{-1}$). 
The ratio of the dispersion 
to the mean value can be taken as a measure of specificity -- for silicon it is 91\%. 
For gold, on the other hand, the calculated \cite{Gottschalk2} binding energies 
vary between 17.5 and 44.2 kJ mol$^{-1}$ with the mean of 30.1$\pm$7.5 kJ mol$^{-1}$ 
which corresponds to the 25\% specificity.

Here, we consider the case of ZnO -- 
it is a well studied semiconductor \cite{jpcmwang} 
with potential applications in biosensing \cite{thai,sarikaya}.
ZnO-based quantum dots can act as inorganic fluorescent probes \cite{wang,gu,godlewski}, 
though problems with biocompatibility, toxicity, sufficient brightness of fluorescence, 
and water-solubility \cite{zhuang} remain to be worked out.
For instance the issue of a relatively easy solubility can be dealt with either 
by coating the dots or by envisioning single-use devices for detection of, say,
enzymatic reactions in blood.
Another application involves self-assembly of biomaterials 
with peptide-linked ZnO nanoparticles \cite{togashi,tomczak}.

We consider the case of four surface cuts of ZnO immersed in water.
We find that AAs split into binding and not-binding in a way that 
depends on the  surface cut. If they can adsorb, the corresponding
binding energy, $\epsilon$, reach up to
7 kJ mol$^{-1}$ which corresponds to 562 K
and there is a considerable specificity.
The highest of these energies are then comparable to the strength 
of the hydrogen bonds in proteins. 
Variations among AAs for a given surface
are found to exceed even 150\%. 
The specificity in the absence of water is  only of order 19\%, 
but the $\epsilon$ is much stronger -- 
of order 147 kJ mol$^{-1}$.

Our results are obtained through numerical modeling. When choosing a modeling scheme
one may take a solid-state perspective which suggests using the quantum molecular dynamics 
(MD) method, as in the studies of alanine near ZnO \cite{gao} 
or various amino acids (AAs) near gold \cite{Gottschalk1}.
However, this approach is restrictive about the time scales and the size of the system, 
making it hard to include water molecules and conformational changes 
of even small biomolecules. 
Here, we adopt the approach of the protein science 
and use the established all-atom classical MD protocols. 
This approach has been taken recently 
in studies of interaction energies and conformational changes 
of proteins anchored to the polar mica surface \cite{kubiak2,starzyk}.
The usage of the classical MD in systems involving ZnO seems justified 
since the solid is not metallic and the electron delocalization effects are weak.

We use the umbrella sampling method \cite{umbrella0,umbrella} 
to derive the potential, $V(z)$, 
of the mean force (PMF) for single AAs in water solutions 
as a function of the distance $z$ above a surface of ZnO. 
The depth of the lowest negative minimum in $V(z)$,
defines the parameter $\epsilon$. There could also be
positive local energy minima that provide trapping
but $V(z)$ would be higher than at large separations from the surface.
$V(z)$ is optimized over the lateral directions $x$ and $y$
and is averaged statistically.

We then consider MD of a small protein, the tryptophan cage, 
and characterize its binding-unbinding behavior near the ZnO surface.
The weak $\epsilon$'s associated with ZnO means that 
binding is expected to be temporary and implementable in many ways.
In addition, the distortion of an adsorbed protein
is likely to be weak and thus occuring without any loss of biological
functionality. This feature should make ZnO-based devices to be good platforms
for bio-functionalization.

ZnO in its crystalline form under normal conditions has a hexagonal wurtzite-type structure
(see Fig. \ref{unit_cell}) with the lattice constants $a$=3.25 {\AA} and $c$=5.2 {\AA} \cite{rossler}.
Its surface morphology is defined by four commonly found faces:
(0001)-O, (000$\bar{1}$)-Zn, (10$\bar{1}$0) and (11$\bar{2}$0) \cite{henrich}.
Here the symbol O (Zn) indicates that the top layer is made of the oxygen (zinc) atoms.
The first two are called polar since the first layer of atom has a net charge.
The last two are called non-polar (but this does not imply that they are hydrophobic).
For ZnO powders, the non-polar faces represent about 80\% of the exposed surfaces
and the remaining 20\% are mainly polar \cite{lamberti}. 
Our simulations pertain to all of these faces.

\section{Method}

Our all-atom MD simulation employ the GROMACS 4.0.7  package \cite{GROMACS}
with the AMBER-99 force field \cite{duan}. 
The package has been designed to study proteins in bulk water 
and the introduction of the solid surface requires making modifications. The first 
approximation is that the solid is rigid and bulk-like, i.e. not distorted at the surface. 
We envision the space above the surface being filled with water molecules 
-- as described by the TIP3P model \cite{jorgensen} 
-- and with either an AA or a protein. 
The biomolecules and water molecules have partial charges on their atoms. 
They interact with one another electrostatically but also through
the Lennard-Jones like interactions which account for repulsive cores
and dispersive features of the atoms. Similar sets of interactions
involve couplings with the atoms of the solid. The Lennard Jones length, $\sigma_A$, and
energy, $\epsilon_A$, parameters for atoms, $A$, in ZnO have been determined \cite{raymand} 
through the density-functional theory (DFT). 
Following this reference we take 2.128 {\AA} for $\sigma_O$,
and 0.418 kJ mol$^{-1}$ for $\epsilon_O$ for the O atom. 
For the Zn atom we take 1.711 {\AA} for $\sigma_{Zn}$,
and 1.254 kJ mol$^{-1}$ for $\epsilon_{Zn}$. 
Generally, for atoms $A$ interacting with atoms $B$, 
we determine the corresponding parameters through
$\epsilon_{AB}= \sqrt{\epsilon_A \epsilon_B}$ and $\sigma_{AB}=\frac{1}{2}(\sigma_A + \sigma_B)$.

The standard procedure of making the simulations feasible 
is to apply cutoffs for all relevant interactions.
We use the cutoff, $d_c$, of 1.0 nm  combined with 
the gradual switching off of the interactions between $d_c$ and 1.2 nm. 
Water molecules together with the AA 
are placed in a space of size $L_x\times L_y \times L_z$ nm,
where $L_x$ and $L_y$ depend on the surface and are about 3.5 nm each.
In proper runs, $L_z$ is equal to 4 nm.
A reflecting wall for water molecules is placed at $z$=4 nm. 
Above this wall, there is an empty space extending to $z$=12 nm. 
Another reflecting wall, just for the amino acids, is placed at $z$=3.5 nm -- otherwise
a protein may get trapped at the water-vacuum interface. The purpose of this construction
with the vacuum  is to allow for the usage of the periodic boundary conditions 
(in the periodic image, above the empty space there are the surface atoms) 
with the pseudo two-dimensional particle mesh Ewald summation \cite{Essman-Darden}.

The MD simulations are performed using the leap-frog algorithm 
with a time step set of 1 fs. 
Temperature coupling with a Berendsen thermostat is implemented
with a time constant of 0.1 ps at a temperature of 300 K.
Initial velocities are generated according to the Maxwell distribution 
and then the system is energy-minimized by
using the steepest descent algorithm. 
Next, the system is equilibrated through 4 ns of MD with $L_z$ set to 5 nm.
This results in water at the top being depleted 
due to attraction at the bottom and then $L_z$ gets reduced to 4 nm.
Further equilibration continues for 1 ns and only then
the biomolecules and Na$^+$ and Cl$^-$ ions are inserted adiabatically.
The concentration of the ions corresponds to the physiological 150 mM:
4 ions of Na$^+$ and 4 ions of Cl$^-$.
If a biomolecule has a net charge due to the side groups, 
extra ions are inserted to neutralize the net charge.

An AA in a protein is in the unionized and not zwitterionic form.
Thus in studies of a single but unionized-like AA we attach caps 
to both sides of the molecule 
(the acetyl and N-methylamide groups to the N- and C-terminus respectively) 
that eliminate the terminal charges and mimic the presence of a peptide chain
(see, e.g. \cite{Gottschalk2}).
Histidine is considered in its three possible protonation states: 
HIE (H on the $\epsilon$ N atom), HID (H on the $\delta$ N atom)
and positively charged HIP (H on both $\epsilon$ and $\delta$ N atoms).
At the assumed value of pH of 7, 
all three forms are present in equilibrium.
The VMD software is used for viewing and analyzing the MD results.

The PMF is an effective potential that yields the average force \cite{Kirkwood}.
We use the umbrella sampling \cite{umbrella0,umbrella1} 
to determine it for the center of mass (CM) of the capped AAs. 
In the first stage of the method, 
a set of initial conformations for representative values of $z$ 
is generated by pulling the CM of the AA along the $z$-axis. 
Pulling is implemented through a "dummy particle" which moves
towards the surface with a constant speed of 1 nm ns$^{-1}$ from $z$=2 nm to $z$=0
and drags the CM by a harmonic force 
(the spring constant, $k$, is 5000 kJ mol$^{-1}$ nm$^{-2}$). 
The lateral motion is not constrained.
The conformations are scanned every 0.1 ps in order to save at
least 30 of them with the CM within each of the interval of 0.05 nm
(0.02  nm in vacuum, combined with a larger $k$). 
In this way, 
about 35 conformations (135 in vacuum) are collected for each AA.
They are used in the second stage for 5 ns of further runs each 
(3 ns correspond to equilibration) 
in which the $z$-location of the pulling particle is fixed and the
CM moves within a sampling window of width $\Delta z$. 
The distribution of the resulting vertical locations of CM (see Fig. 16  
in Electronic Supplementary Information -- ESI$\dag$) 
in the window has a maximum where the harmonic pull balances  all forces 
acting on AA (without the caps) in the $z$ direction. 
This force is averaged over time and distance within each window
and integrated over $z$ to get the PMF.

\section{Results}

The plots of $V(z)$ for the CM of all AAs at the four interfaces can be found in ESI$\dag$.
Here, as an illustration, we consider the case of glycine. 
Figure \ref{glycine} shows that the PMF has the form of a potential well 
with a strong repulsion at small values of $z$ 
combined with ondulations in the attraction part 
which are observed when moving away from the surface. 
We describe the potential well through its depth, $\epsilon$, 
and the vertical distance, $\sigma$, of the deepest minimum. 
For the glycine on the (10$\bar{1}$0) surface, 
$\sigma$ is 0.31 nm in vacuum, and 0.47 nm in water.
The corresponding values of $\epsilon$ are 116.24 kJ mol$^{-1}$ 
and 2.57 kJ mol$^{-1}$.
Note a profound role of the solvent: 
its introduction reduces the binding energy by a factor of 45.2
and shifts the optimal binding distance by 0.16 nm away from the surface. 
Tables \ref{tab_eps} and \ref{tab_sig} indicate that similar effects of water
are observed for other AAs. 
The noted significance of the solvent for interactions with ZnO is in
contrast to a recent finding by Pandey et al. \cite{pandey} pertaining
to interactions with graphene. In their model, no electric charges
are associated with the carbon atom of graphene and hence the density profile
of water gets patterned only in a minor way through the van der Waals
interactions. Thus there is no buildup of solvent density in the first
layer. Thus AAs are able to interact with graphene directly,
i.e. nearly as in vacuum.

When looking at the results obtained for (10$\bar{1}$0) in vacuum, 
we note the exceptionally high binding energy of histidine, 
including its unprotonated form (HID). 
The corresponding $\epsilon$ is the highest obtained among the uncharged AAs (see Table \ref{tab_eps}). 
This finding is consistent with the stability of proteinic structures, 
such as the zinc finger proteins \cite{hanas}, 
in which the Zn$^+$ ion is coordinated by two (unprotonated) histidines and two cysteins.
The reason for the proteinic and surface situations being comparable is that
an interior of a protein is typically devoid of water molecules.
Another comparison can be  made with the data on layers of non-capped alanine
placed on the (10$\bar{1}$0) surface of ZnO in the ultra high vacuum apparatus 
and studied by the X-ray photoelectron spectroscopy \cite{gao}. 
It has been determined that alanine binds in vacuum and the coverage is about 0.4 nm thick.
Alanine in its monodentate and bidentate forms has been found to bind 
with the energies of 1.03 and 1.75 eV respectively, 
compared to our 110.98 kJ mol$^{-1}$, i.e. 1.15 eV.
Our system is more similar to the  monodentate form so the agreement is close.
The zwitterionic form has been found not to be stable on ZnO.
However, in the DFT calculation in which it is constrained to stay at the surface, 
the binding energy is predicted to be 1.39 eV \cite{gao}.
These results suggest that our PMF for the capped alanine in vacuum 
has the right order of magnitude.

Table \ref{tab_eps} and Figure \ref{aa_ran} show that the highest
binding energies are comparable to that in a hydrogen bond 
(6.9 kJ mol$^{-1}$ for O-H binding N and 5.0 kJ mol$^{-1}$ for O-H binding O). 
Among the four faces,
glycine exhibits the strongest attraction 
to (000$\bar{1}$)-Zn -- 3.36 kJ mol$^{-1}$ but experiences
repulsion from (0001)-O. 
If one considers all AAs and all four faces, the strongest
attraction is found for tyrosine at the (000$\bar{1}$)-Zn surface.

The reason for the all-changing role of water is that 
it gets adsorbed to the surface in a profiled manner 
in an analogy to what has been observed in the Lennard-Jones fluids 
near attractive walls \cite{cieplak,cieplak1}.
The water number density profiles shown in Figure \ref{wat_den} 
display at least two well defined maxima. 
The presence of the maxima is 
consistent with the DFT calculations for single molecules of water 
placed at distances corresponding to the first layer above the (10$\bar{1}$0) ZnO surface 
where the largest binding affinity has been found. 

Experimental studies of water molecules on metals \cite{carrasco} 
indicate existence of rich variety of structures 
with both H-up and H-down orientations. 
For ZnO, however, we observe the H atoms of the molecules in the first layer 
to prefer orientations that are closer to the surface 
but in a manner that depends on the surface (see Fig. 14 in ESI$\dag$).

The density profiles depend on the type of the surface. 
For (10$\bar{1}$0) the first layer is 0.08 nm closer to the surface than in the case of (0001)-O 
and the gap between the two layers is more articulated. The gap is even more
articulated for the (000$\bar{1}$)-Zn surface.

The first layer fluctuates much less than the second
(in 1 ns only two molecules exchange between the two 
in the case of (11$\bar{2}$0) - no exchange has been detected in other cases)
but they both provide screening from the surface and hinder a closer
approach of a biomolecule to the surface.

\begin{figure}[ht]
\begin{center}
\includegraphics[scale=0.30]{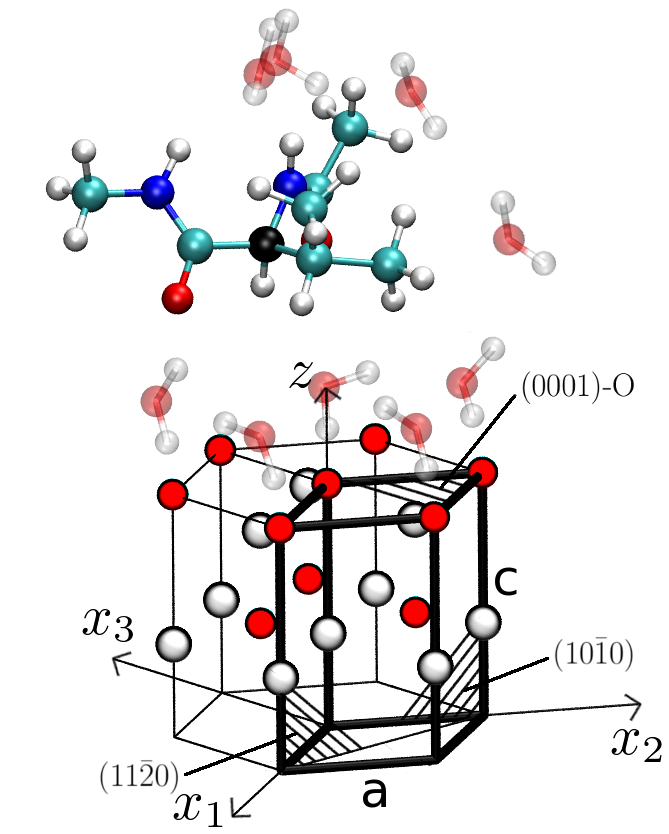}
\caption{Definition of the system. The lower part shows the crystalline structure of the ZnO. 
The $x_1$, $x_2$, $x_3$ and $z$ axes indicate the main directions in the hexagonal system.
The first three pass at the $120^{\circ}$ angle to one another. 
$a$ and $c$ are the lattice constants. 
The O atoms are shown in red and Zn in white within black circles.
The unit cell is highlighted by the bold lines. 
The faces studied are shaded by the stripes.
The (0001)-O face is at the top and (000$\bar{1}$)-Zn at the bottom. 
The upper part shows the optimal conformation for valine 
surrounded by water molecules at the PMF minimum.
The water molecules below valine belong to the first density layer.
One molecule of water shown on the right of the AA belongs to the second layer.
The C$^{\alpha}$ atom is highlighted in black. 
The H, C, and N atoms are in gray, green, and blue respectively.}
\label{unit_cell}
\end{center}
\end{figure}

\begin{figure}[ht]
\begin{center}
\includegraphics[scale=0.3]{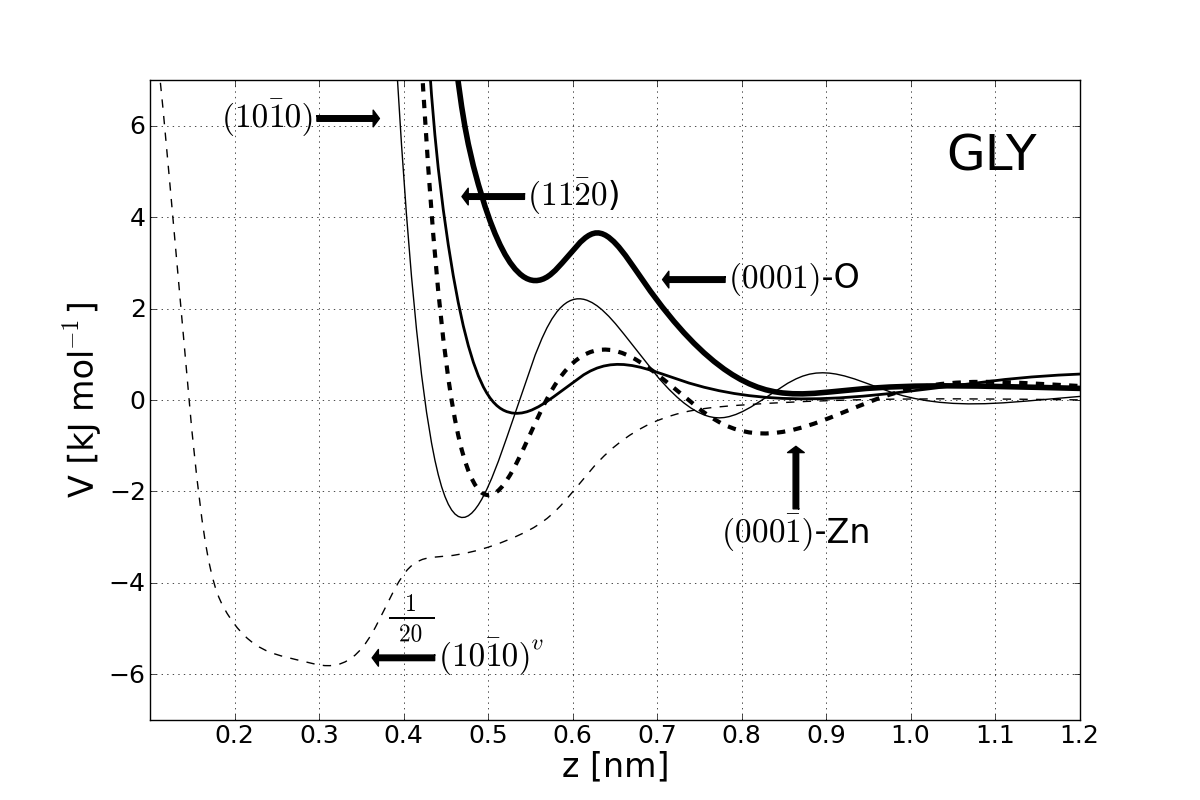}
\caption{PMF for glycine for the indicated surfaces.
All results are in water except for the dashed line 
also marked as (10$\bar{1}$0)$^v$
(the corresponding data is rescaled by the factor of 20.}
\label{glycine}
\end{center}
\end{figure}

\begin{figure}[ht]
\begin{center}
\includegraphics[scale=0.50]{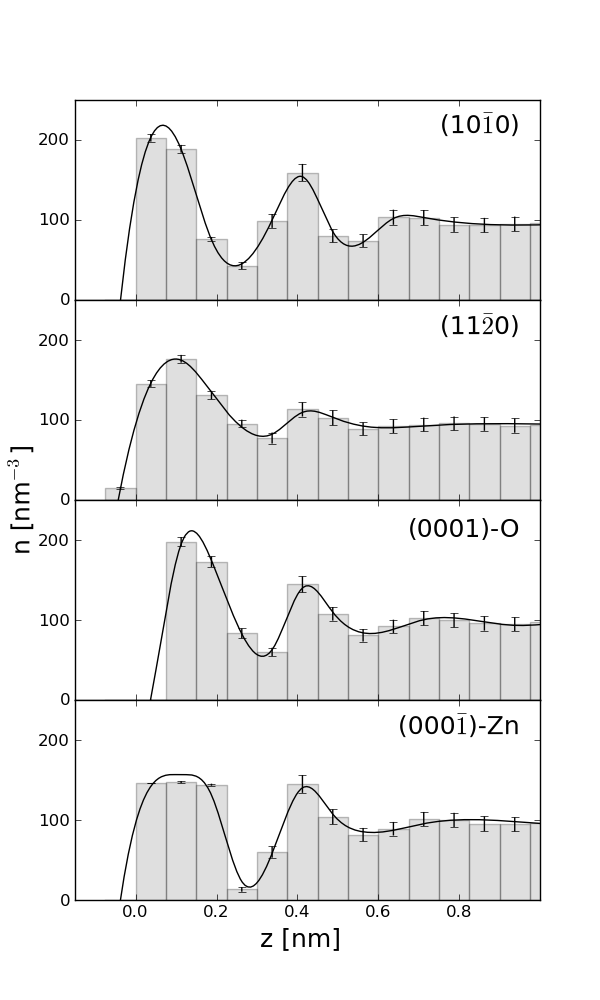}
\caption{The number density profiles, $n$, of water molecules 
above the four faces of ZnO. 
The profiles are shown as a function of $z$. 
They are averaged over the $x-y$ plane and also time-averaged.
In the case of (10$\bar{1}$0), three density maxima are observed.
In the case of (11$\bar{2}0$), 
there is some penetration of water just below the surface.}
 \label{wat_den}
\end{center}
\end{figure}

\begin{figure}[ht]
\begin{center}
\includegraphics[scale=0.25]{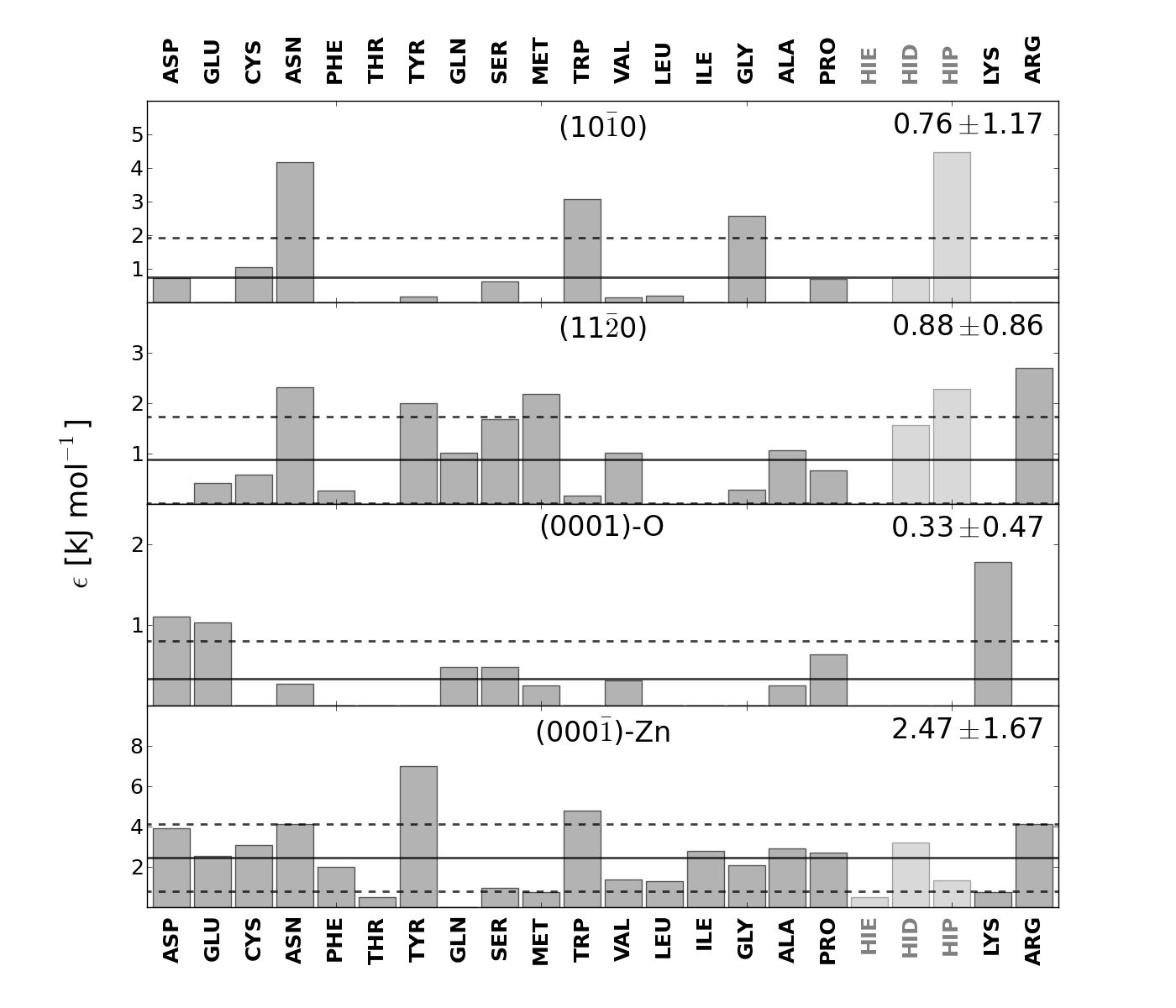}
\caption{Values of $\epsilon$ for AAs in water.
The horizontal solid line shows the average. 
The horizontal dashed lines show the standard deviations.
When calculating the averages, 
the three forms of HIS are considered as a single AA.
} \label{aa_ran}
\end{center}
\end{figure}

An inspection of the average values of $\sigma$'s given in Table \ref{tab_sig}
leads to the conclusion that, at the deepest minimum of $V(z)$, 
the CM of an AA is above the second layer 
 -- for each of the four surfaces.
The corresponding CM of the side group is either between the first two
layers or within the second layer.
As illustrated at the top of Figure \ref{unit_cell} qualitatively, 
the AA displaces water level at the second layer and makes a cavity for itself.
Its interactions with the wall are screened by the first layer 
-- a direct approach to the bare surface is unlikely. 
The motion of the pulling particle has rarely led 
to some penetration of the first layer.
It should be noted that the calculations were performed by adding AAs to water. 
If water was added after the adsorption of an AA in vacuum 
then no dissociation was observed. 
(However, the total energy of the system becomes
 less negative in the latter case.)
The reason for this behavior is that when water is present from the
begining, the density profiling sets in and the formation of the 
first layer prohibits a closer approach of an AA to the surface.

It should be noted that the density profile of water near Au(111) \cite{Gottschalk2} 
is quite distinct compared to the profiles found near ZnO. 
The two are compared in Figure 15 in ESI$\dag$ 
where the case of ZnO is represented  by the (10$\bar{1}$0) face. 
The first density maximum near the Au surface is almost 50\% higher 
and noticeably narrower than near the ZnO. 
The corresponding PMFs have minima right in the middle of the first layer \cite{Gottschalk2}; 
see also \cite{feng_2010}.
This leads to the AAs approaching closer to the surface, breaking the
first layer and binding stronger.
The reason for such a behavior is relate to the details
of the modeling of Au surface \cite{iori_2009}. In particular, 
the effective charges are more than three times smaller 
than in the model of ZnO surface, e.i. 0.3 and -0.3$e$.
These charges are even smaller than in two models of the water molecule, 
SPC and TIP3P, i.e. about 0.4 and -0.8$e$ for H and O atoms respectively.
Furthermore, the opposite charges on gold are merely 0.07 nm apart, 
which makes the net electric field small  but accounts for the
polarizability properties of the metal.
For comparison, Zn and O atoms are separated by about 2 nm
and bind water much stronger making the first layer impenetrable.


Our results for ZnO are qualitatively consistent with related simulations \cite{monti} 
pertaining to ammonium, methane, methanol, methanoate, benzene and guanidinium 
which are analogues of the side chains of lysine, alanine, serine, 
aspartic acid, phenylalanine, and arginine respectively, 
at the aqueous rutile TiO$_2$ (110) interface. The effective
charges used were 2.196$e$ and -1.098 for the Ti and O atoms respectively.
We get energies of the similar order of magnitude,  plots of $V(z)$ also come
with minima both at positive and negative energies, and there are
also repulsive situations. Some analogues tend to bind to the 
first layer of the solvent instead of to the solid.


We now discuss sequences of AA's. 
One example is ZnOBP -- the ZnO binding peptide.
It has been recently discovered \cite{yokoo} 
that ZnOBP suppresses the (0001) growth from zinc hydroxide
so it can be used to grow flattenned ZnO nanoparticles 
-- an approaching Zn atom cannot attach to the cristal directly. 
The "hot spot" in the peptide has turned out to be 
MET-HIS-LYS at its 5-6-7 sequential sites. 
This finding has motivated Togashi et al. \cite{togashi} 
to consider dipeptides MET-HIS and HIS-LYS for the same purpose. 
They work even better than ZnOBP whereas no effect was detected 
for the control dipeptides ALA-ALA, ALA-HIS, and HIS-VAL 
or the monomeric HIS, MET and LYS. 
These findings are surprising when confronted with the values of the binding
energy listed in Table \ref{tab_eps}. For instance, at the (000$\bar{1}$)-Zn 
surface,
$\epsilon$ of both ALA and VAL is larger than of MET or LYS.
This suggests that binding energies of combined objects are not simply additive 
since interactions between AAs lead to conformational changes that may 
affect adsorption. 

This lack of additivity has been shown to arise in peptides near 
the (100) Si surface \cite{Bachmann}: various
sequential arrangements of a fixed set of AAs lead to different binding affinities. 
A similar work \cite{Oren} involved a host of septapeptides near the surface of Pt 
and various degrees of affinities have been measured due to conformational flexibility. 
However, flexibility of proteins is generally reduced compared to short peptides 
so binding affinities may become more dependent on the single AA binding strength. 
None of the binding energies obtained here ensures a permanent AA adsorption 
to ZnO which agrees with the experimental findings on adsorption
of monomeric AAs \cite{togashi}.  We find that the
(000$\bar{1}$)-Zn surface should attract AAs much better than 
(10$\bar{1}$0) or (11$\bar{2}$0) which explains the (0001) face 
suppression by dipeptids.

Tryptophane cage with the structure code 1L2Y and a sequence of 20 AAs 
is a convenient small protein for further studies of the issues of binding to surfaces. 
Our MD simulations suggest that once 1L2Y approaches a ZnO surface through 
diffusion, it may bind temporarily for between 0.5 and 20 ns. 
In altogether four trajectories of 40 ns for each of the surfaces that start 
with the CM at 2 nm, 
we have observed 34 temporary adsorption events: 8 for the (10$\bar{1}$0)
surface, 6 for 11$\bar{2}$0), 7 for (0001)-O, and 13 for (000$\bar{1}$-Zn). 
In each of these, the set of AAs that coupled to the surface was distinct.
In the case of (10$\bar{1}$0), the coupling is usually due to the flexible lysine.
The temporary nature of the attachment signifies 
that at most several of the bonds form 
and each of them is merely about one hydrogen-bond strong. 
Our pilot studies of the PMF for this protein suggest 
a potential with several minima 
with none of them deeper that -3 kJ mol$^{-1}$.

\begin{figure}[ht!]
\begin{center}
\includegraphics[scale=0.25]{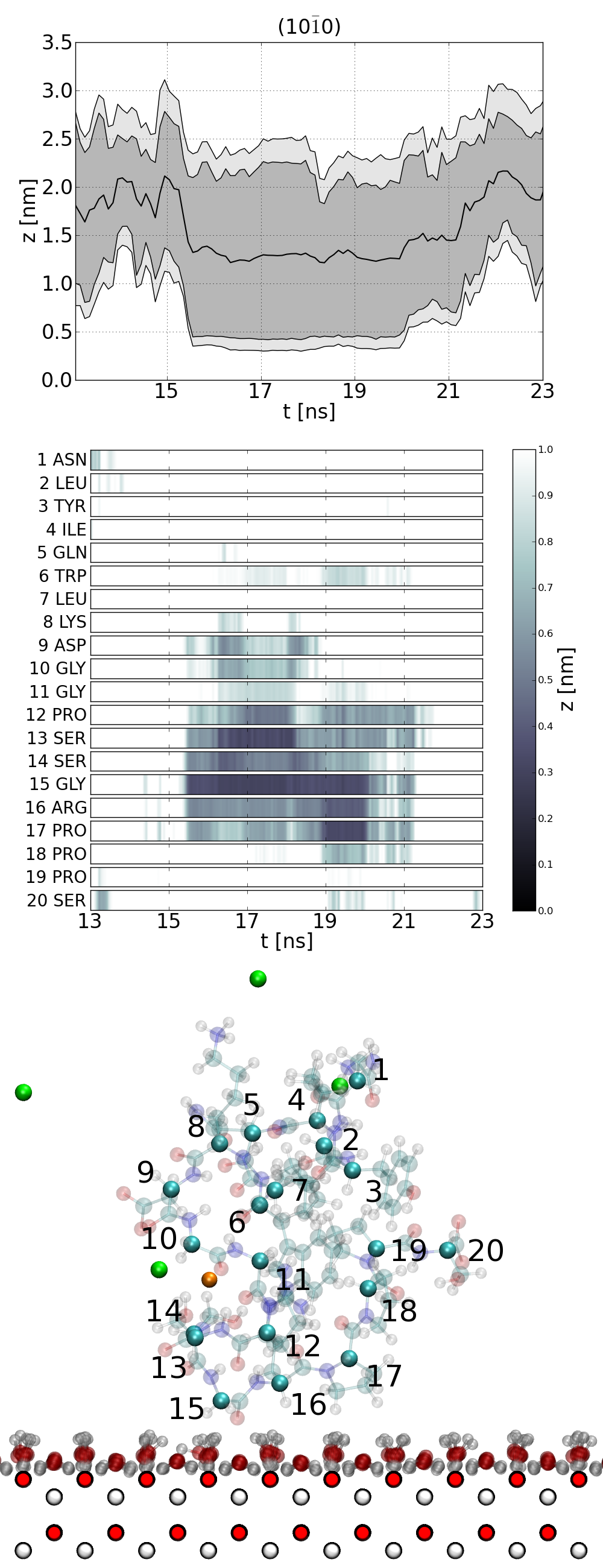}
\caption{The behavior of the tryptophane cage near the (10$\bar{1}$0) surface of ZnO.
The lines in the top panel show, top to bottom, 
the instantaneous vertical positions of: 
the highest atom, the highest CM of an AA, the CM of the whole protein, 
the CM of the lowest AA, the lowest atom. 
The middle panel shows the vertical positions of the lowest atoms 
of all AAs in the protein. 
In the selected time interval we recognize adsorption event between: 18 876 and 19 876  ps. 
The bottom panel shown a snapshot of 1L2Y at time 19 000 ps 
when the protein is adsorbed temporarily. 
Water molecules are shown only in the first layer for clarity. 
The isolated spheres show two of the four ions: 
Cl$^-$ at the top (in green) and Na$^+$ near the surface (in orange).
}
\label{try}
\end{center}
\end{figure}

An example of a binding event shows Figure \ref{try}
for the case of (10$\bar{1}$0). 
The top two panels pertain to a time interval between 13 and 23 ns in a trajectory 
that starts when the lowest atom is around 0.9 nm above the surface.
The protein "touches down" several times and then disengages
before entering the interval we focus on here.
Binding is seen to take place between 15.5 and 20 ns.
The lowest placed AA is 15-GLY and the $z$ coordinate of its lowest atom 
is seen to be nearly fixed at $\sim 0.4$ nm. Between 18.9 and 19.9,
adsorption appears to involve two more AAs: 16--ARG and 17-PRO.
The former has no binding affinity to (10$\bar{1}$0), according to
Table \ref{tab_eps}. It is likely that it has come to the surface
simply because its sequential neighbors got attracted there.

\begin{figure}[ht!]
\begin{center}
\includegraphics[scale=0.25]{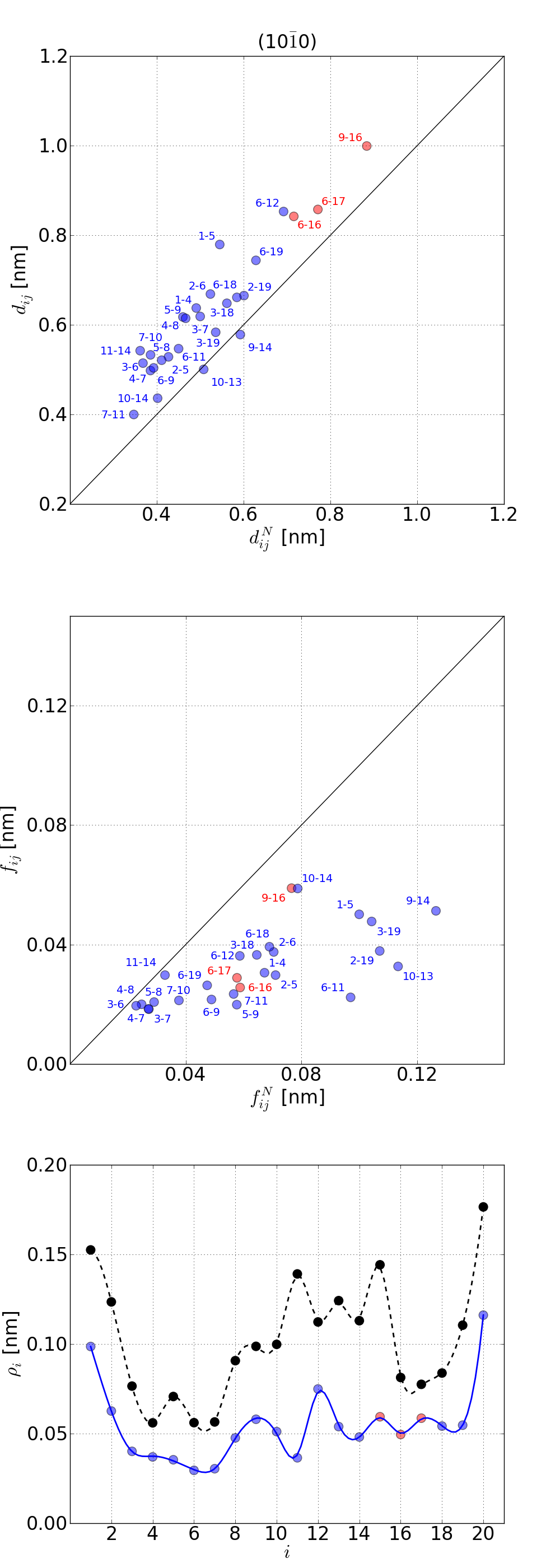}
\caption{Dynamics of 1L2Y during a binding event to the (10$\bar{1}$0) surface.
The top panel plots average distances in the native contacts
in the bound state against the same distances in the bulk water.
The middle panel is similar 
but it relates fluctuations in the length of the native contacts.
The red color is used if at least one AA is involved in the binding.
The bottom panel shows the fluctuations $\rho_i$ for all AAs.
The dashed line corresponds to the data in bulk water 
whereas the solid (blue) line correspond to the binding interval.
The averages are over the time interval between 18 876 and 19 876 ps of the event shown in Figure \ref{try}.
It involves three AAs (15, 16, and 17).
The adsorbed AAs are shown in red.
} 
\label{try_nat}
\end{center}
\end{figure}

The bottom panel shows the conformation of the protein bound at 19 ns.
Its radius of gyration, $R_g$, is equal to (0.706$\pm$0.007) nm, 
which is close to that in bulk water, (0.727$\pm$0.014) nm, 
and yet the adsorbed conformation is seen to be deformed relative to the native state.
The Cartesian components of $R_g$ are found to be equal to
$0.631\pm0.013$, $0.556\pm0.018$, and $0.537\pm0.009$ for the
$x$, $y$, and $z$ components respectively. Thus the surface-normal
direction is the easiest to rotate about.
The $R_g$ does not capture the distortion, 
but average distances, $d_{ij}$, between AAs do. 
This is shown in the top panel of Figure \ref{try_nat} 
where $d_{ij}$ in the native contacts are plotted against their values in bulk water, $d_{ij}^N$. 
The native contacts are defined in a geometric fashion as in ref. \cite{Tsai} 
through the van der Waals volume of the heavy atoms, 
and as used dynamically in Go-like models \cite{JPCM,PLOS}. 
A native contact between AAs $i$ and $j$ is declared to be present 
if their excluded volumes  nearly overlap. 
We observe that all native contacts get longer by between 8 and 38\%.

The second panel shows that also the rms fluctuations, $f_{ij}$ 
in the $d_{ij}$ are distinct from those, $f_{ij}^N$, corresponding to the $d_{ij}^N$.
Most $f_{ij}$'s are smaller than $f_{ij}^N$'s, 
indicating formation of a more rigid structure. 
However, some of the contacts, like 1--4, fluctuate stronger. 
Another way to describe the fluctuational dynamics of a protein is through the RMSF 
- fluctuations in the location of a single AA 
(after subtracting translation of the CM and rotations) and denoted here by $\rho_i$.
The last panel in Figure \ref{try_nat} shows 
that $\rho_i$s in the bound state are smaller than in bulk water.
The suppresion in sites 15 through 17 is by a factor of about 1.3.
Three other examples of adsorption intervals are analyzed in ESI$\dag$. 
Each of them leads to different patterns in the fluctuations 
and different numbers of the AAs are involved in binding.

\section{Discussion}

In summary, we have studied aqueous solutions of AAs near rigid surfaces of ZnO 
by  means of the MD simulations. 
The advantage of this approach over {\it ab initio} methods is that it brings
in many of the ingredients that are important in the studies of adsorption: 
bigger systems, longer times scales, a possibility to consider many 
conformations and include motion of the molecules of water. 
We have found that the layered character of the density profile 
of water near the surfaces 
affects the adsorption process of AAs and is a crucial factor 
that determines the binding strength.

Unlike the situation in vacuum, the values of $\epsilon$ 
are found to depend sensitively
on the precise identity of an amino acids. For instance, the
two protonated forms of histidine (HID and HIP in Table \ref{tab_eps}) 
have distinct values of $\epsilon$. This sensitivity stems from the
fact that the side groups attempt to find an optimal conformation
between the first two layers of water. The layer closer to the solid
is nearly frozen and aligned by the electric field generated by the
solid. It creates an electric field of its own through polarization
and the side groups would tend to find an energy minimum in it.
However, there is also the second layer and the residues also
have to adjust to it. This second layer is mobile and
its interplays with the side groups are transient. The second
layer may repel or attract the residues. Consider the aromatic 
and hydrophobic phenylalanine
above the (000$\bar{1}$)-Zn surface.
The first layer of water here has the H atoms facing the solvent
and forming an almost frozen grid of positive partial charges
with an underlying but laterally shifted plane of negative 
(and twice as big) charges due to O.
The second layer of water alignes the aromatic ring parallel to 
the hydrogenic grid underneath which
brings the partial negative charges of the carbon atoms to the proximity
of the grid. A binding occurs through a proper locking of all ring atoms
into the nonuniform electric field. The binding is enhanced
for tyrosine and tryptophan by also involving their polar groups.
The positively charged histidine has a different ring. This ring is
more tilted relative to the grid in the HIE form whereas 
it is nearly parallel in the HID form.

The optimal PMF energies for interactions of the AAs with ZnO 
are found to be: a) weaker in strength than hydrogen bonds in a protein 
(when all binding situations are considered, for all four surfaces,
$\epsilon = 1.66 \pm 1.47$ kJ mole$^{-1}$),
b) highly specific, 
and c) highly dependent on the selection of the surface.
If one considers all binding situations on all four surfaces
then the $\epsilon=1.63 \pm 1.44$.
This result implies that preparations of bio-sensing or bio-functionalized 
ZnO interfaces require uniformly
ordered crystalline structures that are more costly to prepare.

We predict that a small protein, 1L2Y, should bind to ZnO only intermittently.
It is possible that larger proteins, containing many binding AAs, 
may generate long lasting attachments 
because of more AAs participating in making the coupling 
and larger cancellation of the random forces exerted by the molecules of water 
further away from the surface that might induce detachment.

Analyzing larger proteins will be helped by using coarse grained models
in which the PMFs derived here could be incorporated to describe interactions with ZnO.
Modeling systems with other surfaces in a systematic way is important 
when considering applications in bio-sensing and bio-functionalization.

{\bf Acknowledgments}
Discussions with P. Cieplak, D. Elbaum, J. Grzyb, B. R\'o\.zycki, and M. Sikora 
are appreciated as well as the help of S. Niewieczerza{\l} with the GROMACS code.
This work has been supported by he European Union within European Regional Development Fund, 
through Innovative Economy grant (POIG.01.01.02-00-008/08) 
and by the Polish National Science Centre Grant No. 2011/01/B/ST3/02190.
The local computer resources were financed by the European Regional Development Fund 
under the Operational Programme Innovative Economy NanoFun POIG.02.02.00-00-025/09. 
We appreciate help of A. Koli{\'n}ski and A. Liwo in providing additional computer resources 
such as at the Academic Computer Center in Gda\'nsk.


\clearpage

\begin{table}[h]
\small
\caption{Values of the binding energy $\epsilon$ [kJ mol$^{-1}$] 
between AAs and the four investigated ZnO surfaces. The symbol
-- signifies a non-binding situation (with a possible local
energy minimum at positive energies).
The superscript $v$ signifies results obtained in vacuum.
When calculating the averages and dispersion, as given in the last two
lines, the non-binding AAs are counted as corresponding to $\epsilon=0$
and results corresponding to the various forms of histidine are first
averaged to form one entry.
An expanded version of this Table, containing the values of the
parameters $\sigma$, is provided in ESI.}
\label{tab_eps}
\begin{tabular*}{0.5\textwidth}{@{\extracolsep{\fill}}llllll}
\hline
ZnO & \multicolumn{1}{c}{(10$\bar{1}$0)$^{v}$} & \multicolumn{1}{c}{(10$\bar{1}$0)} 
    & \multicolumn{1}{c}{(11$\bar{2}$0)} & \multicolumn{1}{c}{(0001)-O} & \multicolumn{1}{c}{(000$\bar{1}$)-Zn} \\
\hline
 ASP & 192.14 & 0.72 &   -- & 1.10 & 3.91 \\
 GLU & 197.33 &   -- & 0.42 & 1.03 & 2.56 \\
 CYS & 167.27 & 1.04 & 0.59 &   -- & 3.07 \\
 ASN & 170.38 & 4.17 & 2.31 & 0.27 & 4.12 \\
 PHE & 110.16 &   -- & 0.27 &   -- & 1.99 \\
 THR & 136.39 &   -- &   -- &   -- & 0.51 \\
 TYR & 128.26 & 0.17 & 2.00 &   -- & 7.01 \\
 GLN & 160.66 &   -- & 1.01 & 0.48 &   -- \\
 SER & 140.36 & 0.63 & 1.68 & 0.48 & 0.97 \\
 MET & 123.53 &   -- & 2.19 & 0.25 & 0.77 \\
 TRP & 165.73 & 3.08 & 0.17 &   -- & 4.78 \\
 VAL & 135.23 & 0.16 & 1.02 & 0.31 & 1.39 \\
 LEU & 142.16 & 0.19 &   -- &   -- & 1.28 \\
 ILE & 126.54 &   -- &   -- &   -- & 2.80 \\
 GLY & 116.24 & 2.57 & 0.29 &   -- & 2.08 \\
 ALA & 110.98 &   -- & 1.06 & 0.25 & 2.91 \\
 PRO & 121.15 & 0.70 & 0.66 & 0.64 & 2.72 \\
 HIE & 194.74 &   -- &   -- &   -- & 0.52 \\
 HID & 202.98 & 0.74 & 1.56 &   -- & 3.19 \\
 HIP & 181.40 & 4.47 & 2.28 &   -- & 1.34 \\
 LYS & 169.68 &   -- &   -- & 1.78 & 0.74 \\
 ARG & 126.46 &   -- & 2.70 &   -- & 4.14 \\
average & \textbf{146.68} & \textbf{0.76} & \textbf{0.88} & \textbf{0.33} & \textbf{2.47} \\
dispersion & \textbf{27.41} & \textbf{1.17} & \textbf{0.86} & \textbf{0.47} & \textbf{1.67} \\
\hline
\end{tabular*}
\end{table}

\begin{table}[h]
\small
\caption{Average values of the binding energy $\epsilon$ [kJ mol$^{-1}$]
and the bond length $\sigma$ [nm] 
as measured between the center of mass of an AA and the surface. 
Here, non-binding situations do not contribute to the averages.
Binding forms of histidine count as one average entry.
The superscript $v$ signifies results obtained in vacuum.} 
\label{tab_sig}
\begin{tabular*}{0.5\textwidth}{@{\extracolsep{\fill}}llllll}
\hline
ZnO & \multicolumn{1}{c}{(10$\bar{1}$0)$^{v}$} & \multicolumn{1}{c}{(10$\bar{1}$0)} 
    & \multicolumn{1}{c}{(11$\bar{2}$0)} & \multicolumn{1}{c}{(0001)-O} & \multicolumn{1}{c}{(000$\bar{1}$)-Zn} \\
\hline
average $\epsilon$ & \textbf{146.68} & \textbf{1.46} & \textbf{1.22} & \textbf{0.66} & \textbf{2.60} \\
dispersion in $\epsilon$ & \textbf{27.41} & \textbf{1.33} & \textbf{0.81} & \textbf{0.48} & \textbf{1.61} \\

average $\sigma$ & \textbf{0.28} & \textbf{0.67} & \textbf{0.65} & \textbf{0.84} & \textbf{0.56} \\
dispersion in $\sigma$ & \textbf{0.02} & \textbf{0.19} & \textbf{0.13} & \textbf{0.15} & \textbf{0.05} \\
\hline
\end{tabular*}
\end{table}

\clearpage

\clearpage

{\LARGE Electronic Supplementary Information}

\vspace{0.5cm}

\underline{{\bf Electrostatic potential near ZnO surfaces}}
\vspace*{0.5cm}

The simulations described in the paper can be performed in two equivalent ways.
One is to introduce a cutoff in all Coulombic interactions
and then using the periodic boundary conditions 
with the particle mesh Ewald summation \cite{Essman-Darden}. 
The other is to treat the static electric field of the semi-infinite solid 
as external to the molecular interactions 
and then use the Ewald summation only for the "internal" interactions.
Here, we discuss the properties of the electrostatic potential, $\phi$, 
generated by ZnO that could be used in the second way. 
The potential is non-uniform very near the surface 
but it becomes practically uniform at a sufficiently large elevation. 
It is interesting to find out how does the transition between the two regimes take places.

\begin{figure}[h!]
\begin{center}
\includegraphics[scale=0.40]{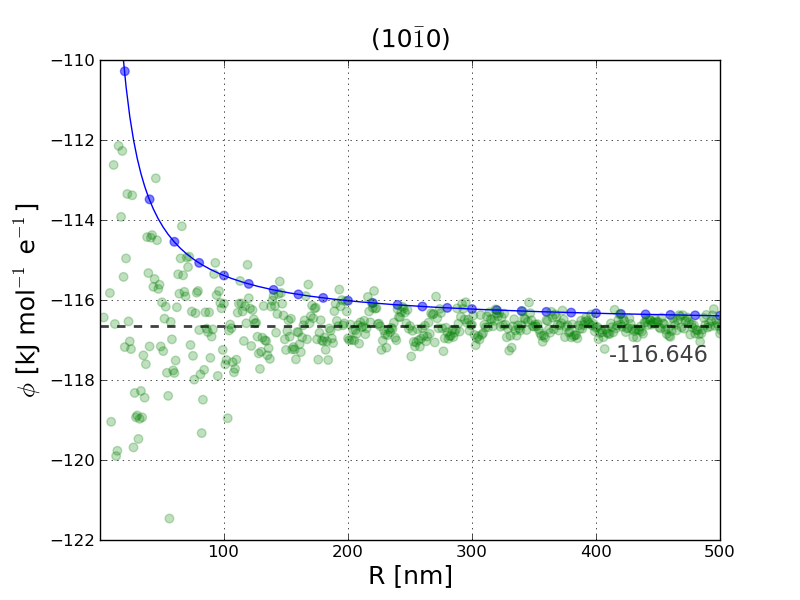}
\caption{$\phi(R)$ for $x$=0.0, $y$=0.0 and $z$=0.2 position above the (10$\bar{1}$0)surface of ZnO.
The dots represent the data obtained: 
the blue circles are for the cylindrical cutouts of radius $R$
and green for the spherical cutouts corresponding to radius $R$.
The blue solid line represents a fit to the cylindrical data points 
whereas the dashed line represents the asymptotic potential obtained 
for the spherical cutouts.}
\label{vr}
\end{center}
\end{figure}

We construct the surfaces by making planar cuts in the bulk solid 
generated by translating the unit cell.
The $\phi$ potential is determined at sites on a grid spaced by $d=$0.01 nm
extending 5 nm above and 0.5 nm below the surface.
The solid is represented by a slab of layers of the unit cells.
For every grid point, we consider a vertical cylinder with radius $R$
and the axis going through the grid point. 
The cylinder contains an integer number of the unit cells 
so its sides are rough. 
The atoms within this cylinder contribute to the Coulombic sum
$C(x,y,z) =f \sum_{i}\frac{q_i}{r_i}$, 
where $r_{i}$ is the distance from the grid site to the $i$'th atom, 
$q_i$ is its charge 
and $f$=138.935485 kJ mol$^{-1}$ nm $e^{-2}$ is the conversion factor. 
The effective charges are -1.026$e$ and 1.026$e$ for the O and Zn atoms respectively, 
as obtained through the Mulliken analysis \cite{raymand}. 
The sums are calculated for a set of values of $R$ up to 500 nm 
and then extrapolated to an infinite $R$ 
by fitting to $\frac{R_0}{R}+\phi(x,y,z)$  ($R_0$ is another fitting constant) 
-- see Figure \ref{vr} for $z$=0.2 nm and $x$=$y$=0.0, i.e. at the O atom.
The values of $\phi(x,y,z)$ are stored. 
The electric field is obtained through discretized differentiation. 
For instance, its $z$-component is given by
$ E_{z}(xyz)=[\phi(x,y,(z-d))-\phi(x,y,(z+d))/2d $. 
The values of $\phi$ and $\vec{E}$ away from the grid points 
can be obtained through interpolation involving the 8 nearest grid nodes.

The electrostatic potential due to the surface has a lateral structure
that depends on $z$. 
Above $z\approx$0.4 nm the electric field becomes uniform 
and close to zero for each surface. 
Modulations in the density of water persist above 0.4 nm 
as they are induced by the field at lower elevations.
It is sufficient to take two layers of cells in the slab 
-- the slab is about 0.5 nm wide. 
Combining, say, 10 slabs one atop another either does not change the $\phi$ or, 
in the case of the polar surfaces, merely adds a constant.

An alternative way to determine $\phi$ is by considering larger and larger spheres 
of radius $R$ (up to 1 $\mu$m) that are cut out in the bulk solid 
and are centered at the grid point. 
The spheres contain same numbers of the Zn and O atoms, but fractional numbers of unit cells. 
Therefore, the nature of the net polarization keeps oscillating 
which results in the oscillatory nature of the  potential. 
Extrapolation to an infinite sphere is, however, 
consistent with the slab results (see Figure \ref{vr}).

\begin{figure}[h!]
\begin{center}
\includegraphics[scale=0.40]{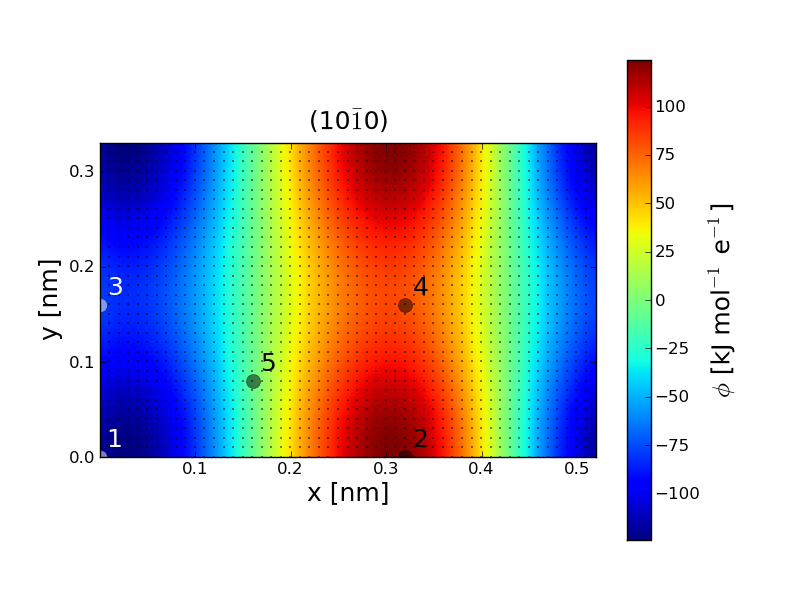}
\caption{$\phi(xy)$ for $z$=0.2 nm above the ($10\bar{1}0$) surface.}
\label{vxy_1010}
\end{center}
\end{figure}

\begin{figure}[h!]
\begin{center}
\includegraphics[scale=0.40]{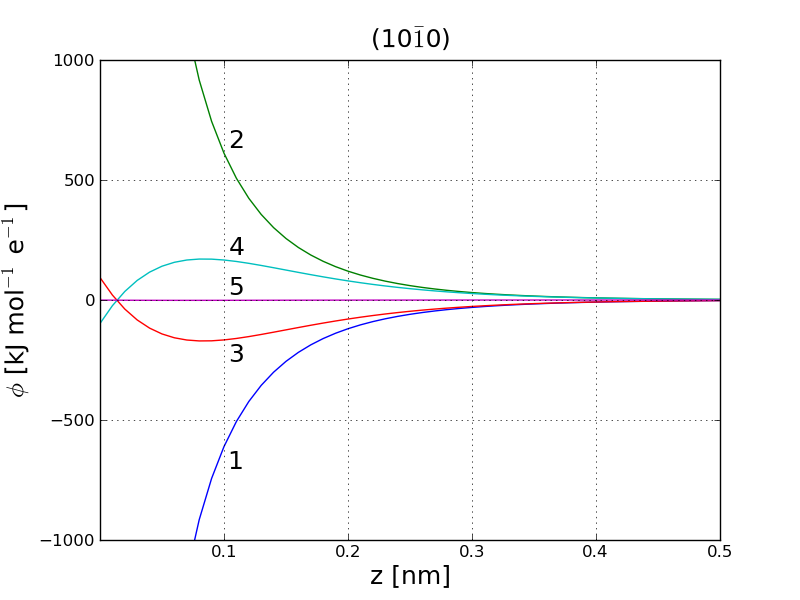}
\caption{$\phi(z)$ for selected $xy$ positions above the ($10\bar{1}0$) surface
as indicated in Figure \ref{vxy_1010}.} 
\label{vz_1010}
\end{center}
\end{figure}

\begin{figure}[h!]
\begin{center}
\includegraphics[scale=0.40]{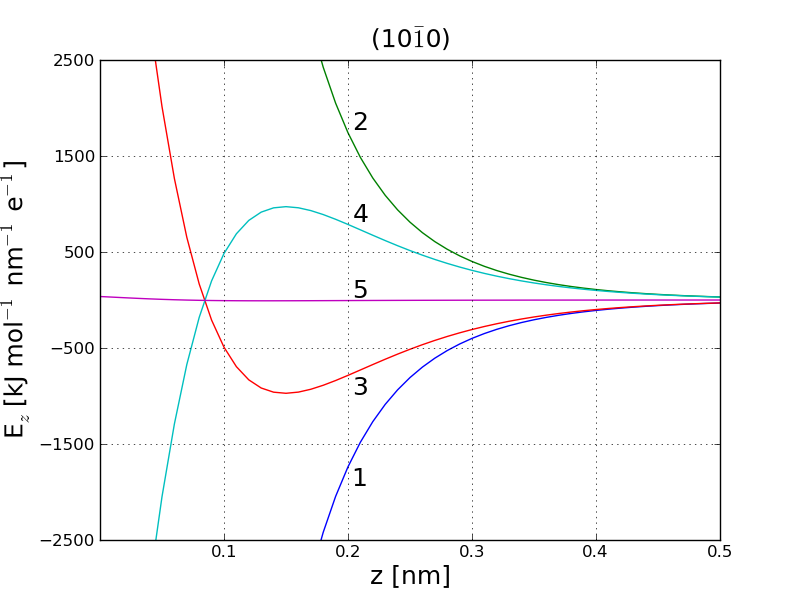}
\caption{E$_{z}$ for selected $xy$ positions above the ($10\bar{1}0$) surface,
as indicated in Figure \ref{vxy_1010}.} 
\label{fz_1010}
\end{center}
\end{figure}

Figure \ref{vxy_1010} shows $\phi$ in the $xy$ plane 
at $z$=0.2 nm above the ($10\bar{1}0$) surface of ZnO. 
Figure \ref{vz_1010} shows $\phi$ as a function of $z$
for selected locations (1 through 5) that are indicated in Figure \ref{vr}.
For instance, 1 is at the origin, i.e. above the O atom.
2 is above the Zn atom. 3 and 4 are above O and Zn atoms
as for sites 1 and 2, but these atoms are located deeper 
-- under the top plane of the surface. 
5 is in the middle.
Figure \ref{fz_1010} shows the corresponding values of the $z$-component of the electric field.

\begin{figure}[h!]
\begin{center}
\includegraphics[scale=0.40]{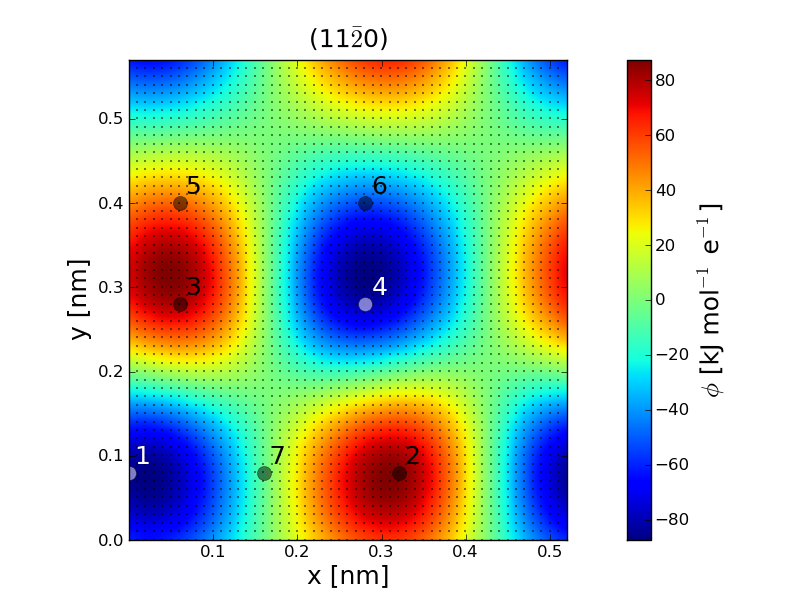}
\caption{$\phi(xy)$ for $z$=0.2 nm above the ($11\bar{2}0$) surface.}
\label{vxy_1120}
\end{center}
\end{figure}

\begin{figure}[h!]
\begin{center}
\includegraphics[scale=0.40]{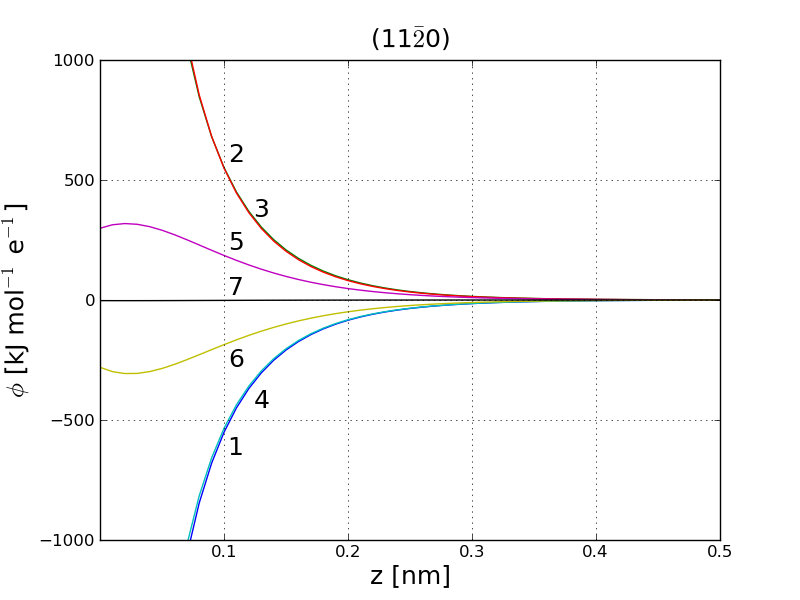}
\caption{$\phi(z)$ for selected $xy$ positions above the (11$\bar{2}$0) surface
as indicated in Figure \ref{vxy_1120}.}
\label{vz_1120}
\end{center}
\end{figure}

\begin{figure}[h!]
\begin{center}
\includegraphics[scale=0.40]{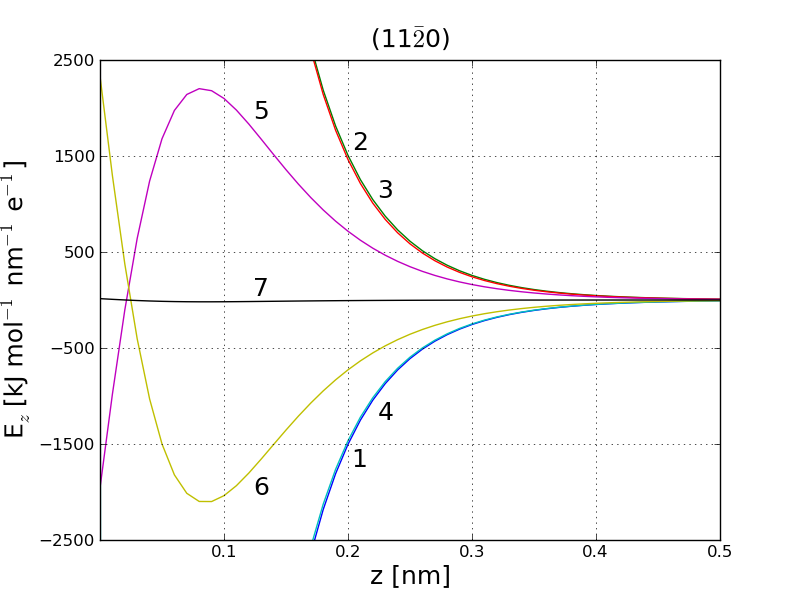}
\caption{E$_{z}$ for selected $xy$ positions above the ($11\bar{2}0$) surface,
as indicated in Figure \ref{vxy_1120}.}
\label{fz_1120}
\end{center}
\end{figure}

\begin{figure}[h!]
\begin{center}
\includegraphics[scale=0.40]{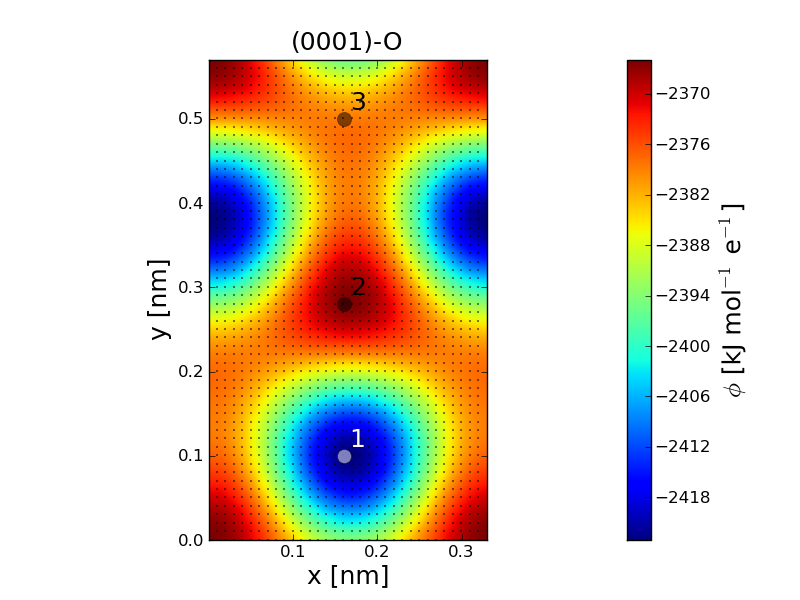}
\caption{$\phi(xy)$ for $z$=0.2 nm above the (0001)-O surface.}
\label{vxy_0001-o}
\end{center}
\end{figure}

\begin{figure}[h!]
\begin{center}
\includegraphics[scale=0.40]{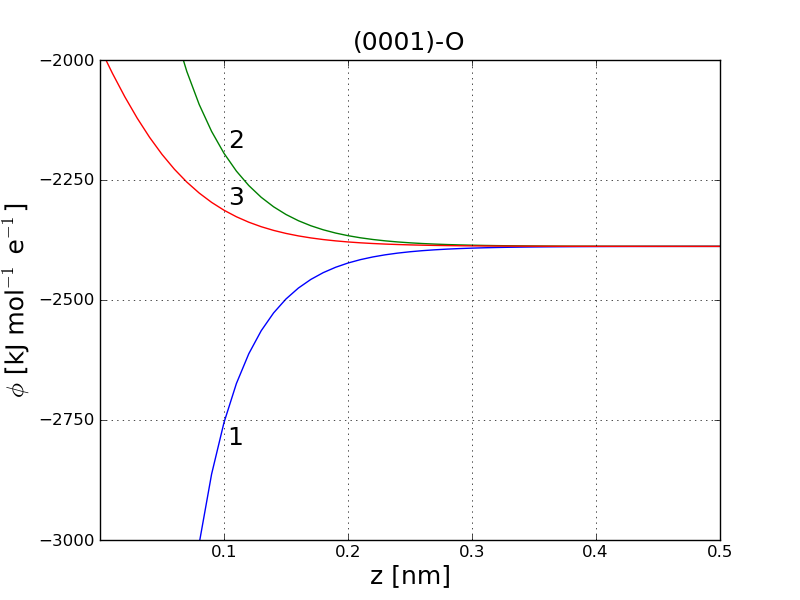}
\caption{$\phi(z)$ for selected $xy$ positions above the (0001)-O surface
as indicated in Figure \ref{vxy_0001-o}.}
\label{vz_0001-o}
\end{center}
\end{figure}

\begin{figure}[h!]
\begin{center}
\includegraphics[scale=0.40]{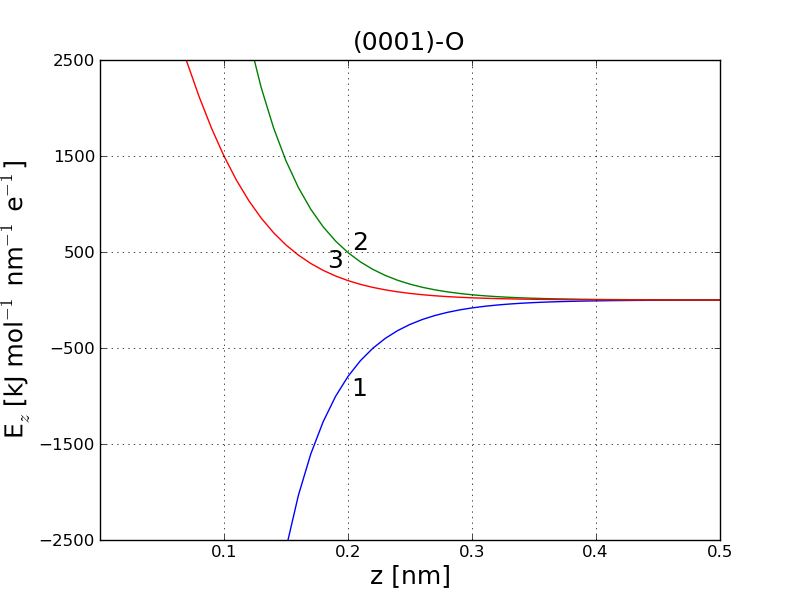}
\caption{E$_{z}$ for selected $xy$ positions above the (0001)-O surface,
as indicated in Figure \ref{vxy_0001-o}.}
\label{fz_0001-o}
\end{center}
\end{figure}

\begin{figure}[h!]
\begin{center}
\includegraphics[scale=0.40]{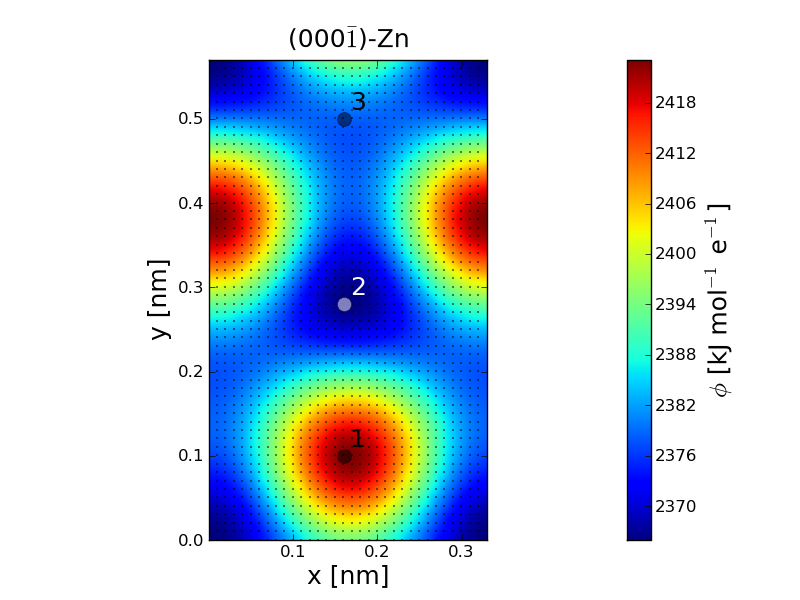}
\caption{$\phi(xy)$ for $z$=0.2 nm above the (000$\bar{1}$)-Zn surface.}
\label{vxy_0001-zn}
\end{center}
\end{figure}

\begin{figure}[h!]
\begin{center}
\includegraphics[scale=0.40]{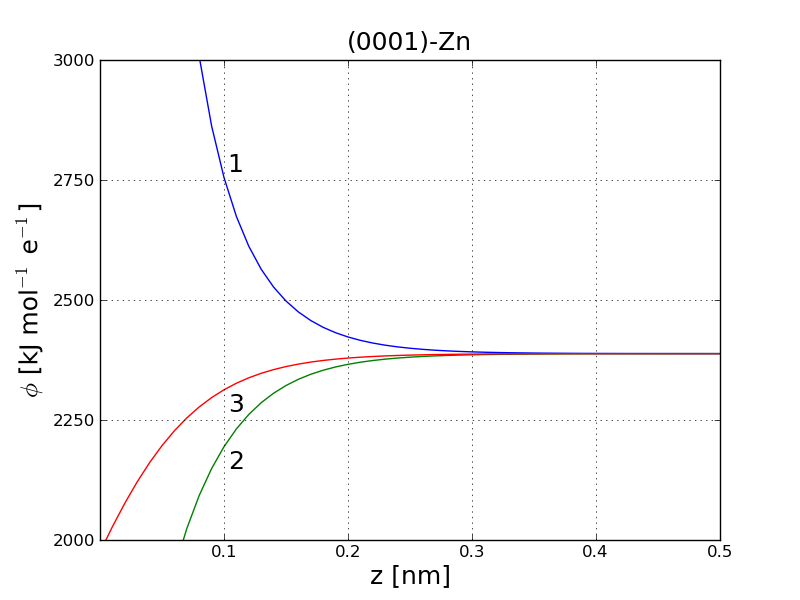}
\caption{$\phi(z)$ for selected $xy$ positions above the (000$\bar{1}$)-Zn surface
as indicated in Figure \ref{vxy_0001-zn}.}
\label{vz_0001-zn}
\end{center}
\end{figure}

\begin{figure}[h!]
\begin{center}
\includegraphics[scale=0.40]{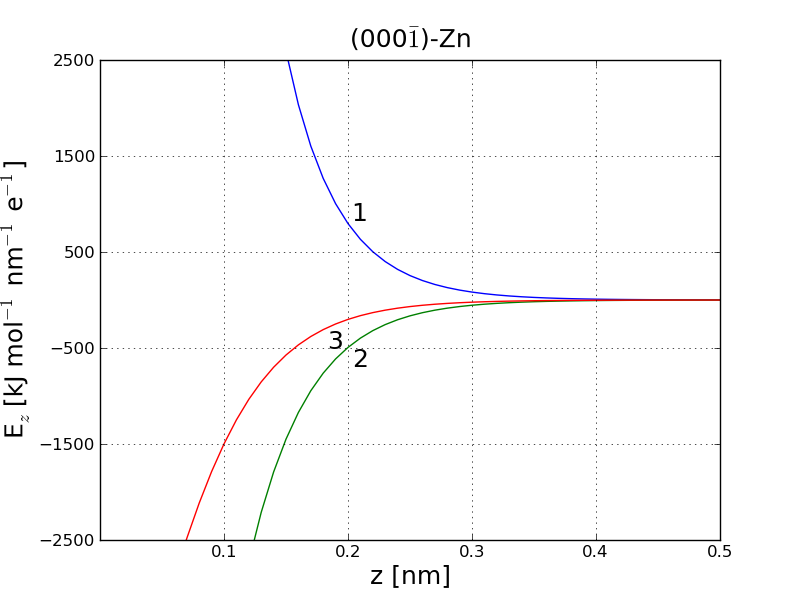}
\caption{E$_{z}$ for selected $xy$ positions above the (000$\bar{1}$)-Zn surface,
as indicated in Figure \ref{vxy_0001-zn}.}
\label{fz_0001-zn}
\end{center}
\end{figure}

The next three figures, \ref{vxy_1120}, \ref{vz_1120} and \ref{fz_1120}, 
are similar but they refer to surface 11$\bar{2}$0.
Figures  \ref{vxy_0001-o}, \ref{vz_0001-o}, and \ref{fz_0001-o} 
refer to the polar surface (0001-O) 
and figures \ref{vxy_0001-zn}, \ref{vz_0001-zn}, and \ref{fz_0001-zn} to the polar surface (000$\bar{1}$-Zn).

\clearpage

\clearpage

\underline{{\bf Structure of water at the solid-water interface}}
\vspace*{0.5cm}

\begin{figure}[ht]
\begin{center}
\includegraphics[scale=0.15]{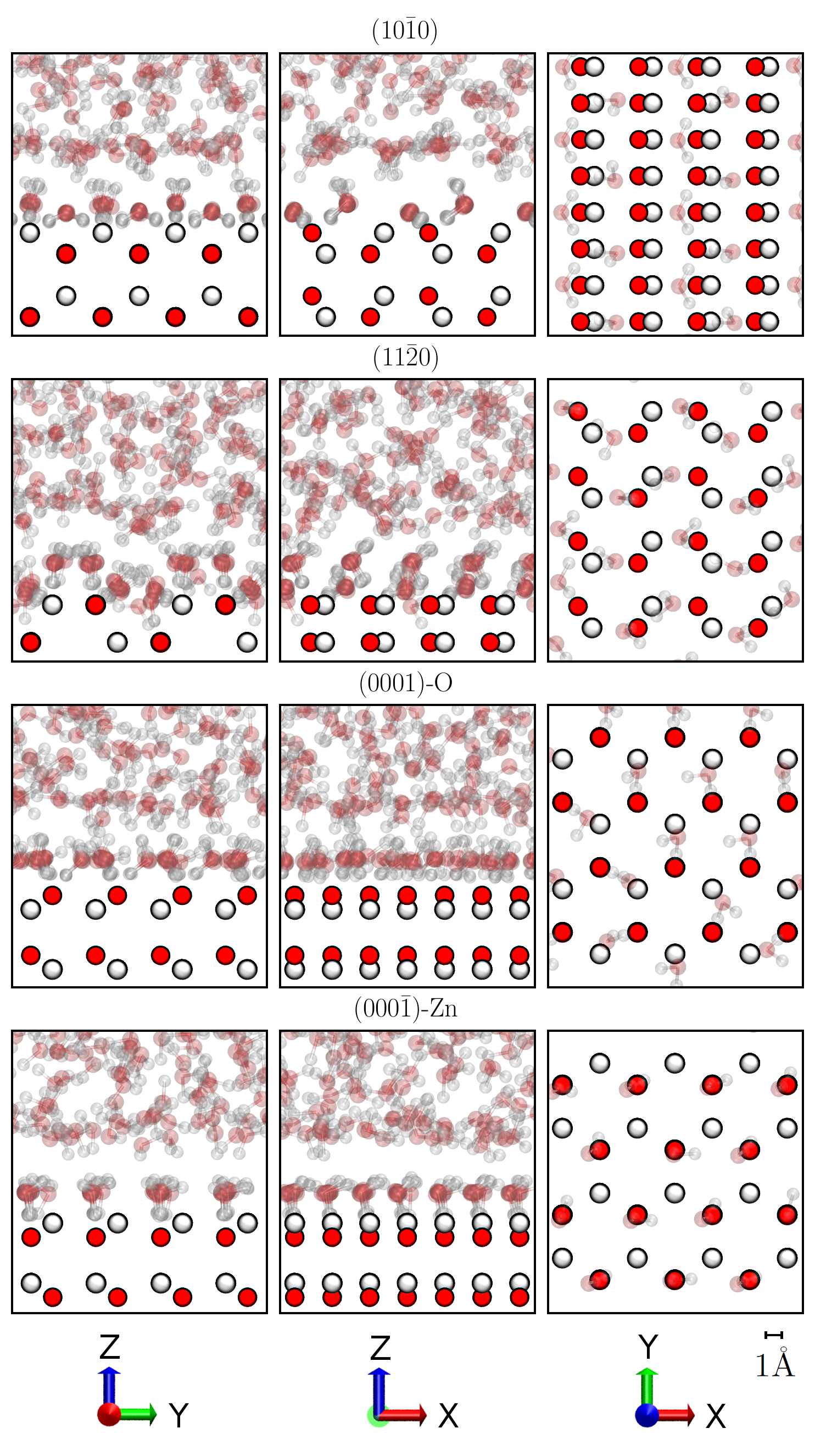}
\caption{Snapshots of the solid-water interface 
for the four surfaces and at various projections. 
For instance, the panels on the right show the top views. 
The larger and stronger symbols show the solid atoms: 
the O atoms are in red and the Zn atoms in white. 
The fainter symbols correspond to the molecules of water 
(in the panels on the right -- from the first layer). 
The red color is again for O and the gray symbols for the H atoms.
Notice, that the H atoms in the first layer tend to face the underlying surface
but in a manner that depend on the surface.} 
\label{zno_wat}
\end{center}
\end{figure}

\begin{figure}[ht]
\begin{center}
\includegraphics[scale=0.45]{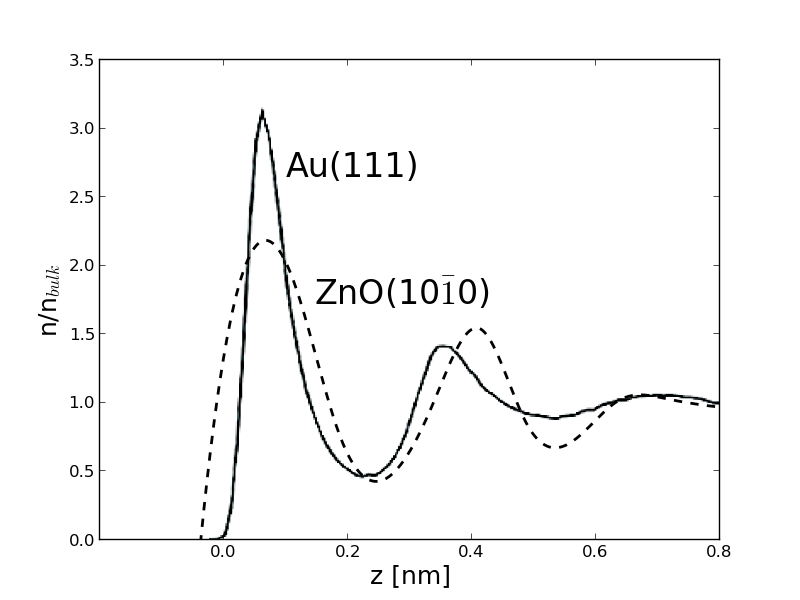}
\caption{
The solid line shows the density profile of water near Au(111)
as obtained in ref. \cite{Gottschalk2}. The density is normalized to 
its bulk value. 
The dashed line is the normalized profile obtained here 
for water molecules near the 10$\bar{1}$0 surface of ZnO.
The location of the first maximum corresponding to the situation with Au(111)
is adjusted to coincide with the first maximum for the ZnO problem.}
\label{auprofile}
\end{center}
\end{figure}

Figure \ref{zno_wat} shows snapshots of the solid-water interface
that illustrate the architecture of the solid 
and the nature of layering of water molecules for the four surfaces 
and at various projections. 
Figure \ref{auprofile} shows the time averaged profile 
of the number density of water molecules near ZnO(10$\bar{1}$0),
as taken from Figure 3
in the main paper, and compares it to the profile near Au(111). 
The profiles are normalized so that the bulk value corresponds to 1.

\clearpage

\vspace*{0.5cm}

\underline{{\bf The umbrella sampling method}}

\begin{figure}[ht]
\begin{center}
\includegraphics[scale=0.31]{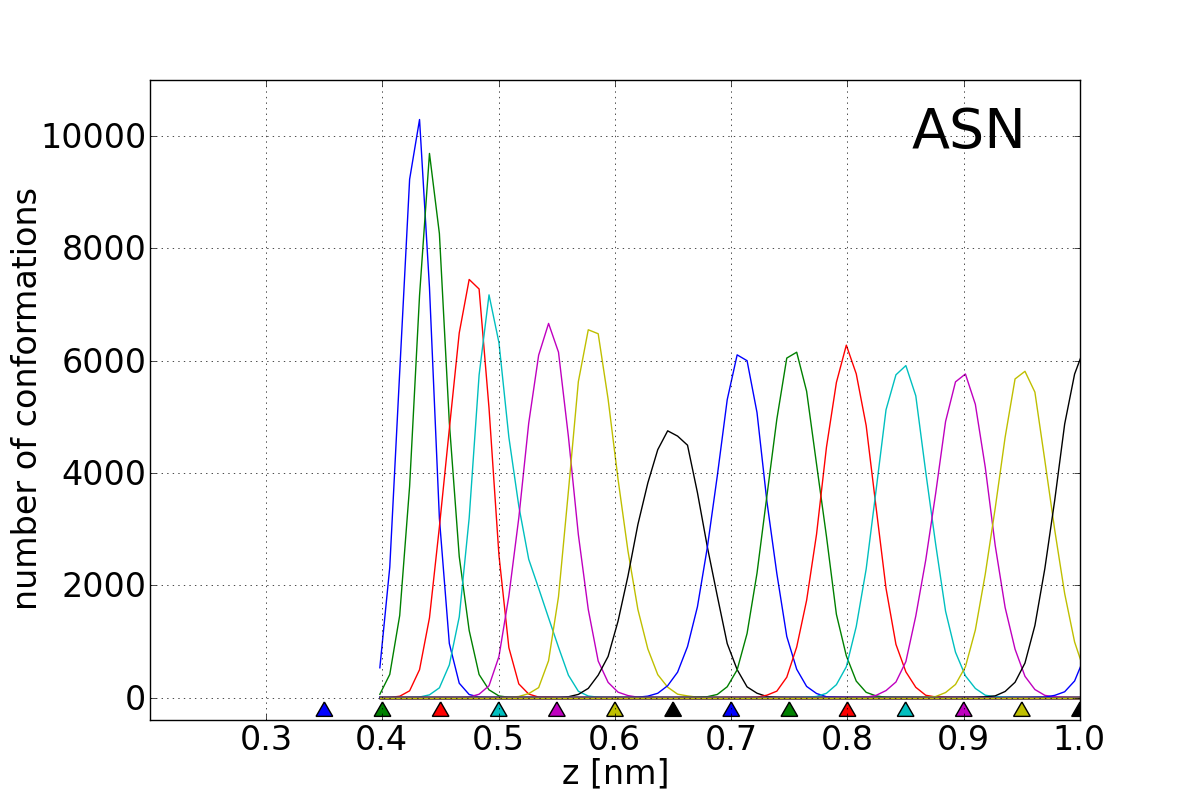}
\caption{Histogram of the number of conformations of asparagine 
above the 10$\bar{1}$0 surface of ZnO in water 
as obtained through the umbrella sampling method.
The different colors correspond to various simulation windows.
In each simulation, the AA is restrained by the umbrella biasing potential 
to different selected values of $z$ that are marked as triangles
n a color corresponding to the simulation window.
For instance, the maximum of the distribution for $z$=0.7 nm
is close to the set value,
indicating a weak impact of the surface.
Between 0.5 and 0.6 nm, the maxima are shifted toward the surface
due to the attraction. However, below 0.5 nm, the maxima are
shifted away from the surface due to the impact of the layers of water.
Generally, a large shift away from the set value comes with
a narrower and taller distribution. A wide distribution, as for $z$=0.65 nm,
suggests that  the AA is attracted by the surface weakly.}
\label{histogram}
\end{center}
\end{figure}

\clearpage

\onecolumn
\underline{{\bf Summary of the results on the binding parameters}}

\vspace*{0.5cm}

Table 1 
provides an expanded version of Table 1 in the
main text. In addition to items listed there it also gives values of the
parameter $\sigma$ for individual amino acids.

\begin{table}[ht]
\label{suptable}
\begin{center}
\begin{tabular}{|c|c|c|c|c|c|c|c|c|c|c|}
\hline
ZnO & \multicolumn{2}{|c|}{(10$\bar{1}$0)$^{v}$} & \multicolumn{2}{|c|}{(10$\bar{1}$0)} 
    & \multicolumn{2}{|c|}{(11$\bar{2}$0)} & \multicolumn{2}{|c|}{(0001)-O} & \multicolumn{2}{|c|}{(000$\bar{1}$)-Zn}  \\
\hline
& \textbf{$\sigma$} & \textbf{$\epsilon$} & \textbf{$\sigma$} & \textbf{$\epsilon$} & \textbf{$\sigma$} & \textbf{$\epsilon$} 
& \textbf{$\sigma$} & \textbf{$\epsilon$} & \textbf{$\sigma$} & \textbf{$\epsilon$} \\
\hline
 ASP & 0.29 & 192.14 & 0.55 & 0.72 &   -- &   -- & 0.56 & 1.10 & 0.44 & 3.91 \\
 GLU & 0.28 & 197.33 &   -- &   -- & 0.91 & 0.42 & 0.61 & 1.03 & 0.52 & 2.56 \\
 CYS & 0.28 & 167.27 & 0.51 & 1.04 & 0.88 & 0.59 &   -- &   -- & 0.56 & 3.07 \\
 ASN & 0.23 & 170.38 & 0.54 & 4.17 & 0.59 & 2.31 & 0.87 & 0.27 & 0.58 & 4.12 \\
 PHE & 0.29 & 110.16 &   -- &   -- & 0.92 & 0.27 &   -- &   -- & 0.55 & 1.99 \\
 THR & 0.31 & 136.39 &   -- &   -- &   -- &   -- &   -- &   -- & 0.58 & 0.51 \\
 TYR & 0.29 & 128.26 & 0.90 & 0.17 & 0.63 & 2.00 &   -- &   -- & 0.54 & 7.01 \\
 GLN & 0.27 & 160.66 &   -- &   -- & 0.63 & 1.01 & 0.95 & 0.48 &   -- &   -- \\
 SER & 0.28 & 140.36 & 0.49 & 0.63 & 0.53 & 1.68 & 0.87 & 0.48 & 0.55 & 0.97 \\
 MET & 0.26 & 123.53 &   -- &   -- & 0.55 & 2.19 & 0.89 & 0.25 & 0.54 & 0.77 \\
 TRP & 0.33 & 165.73 & 0.63 & 3.08 & 0.61 & 0.17 &   -- &   -- & 0.53 & 4.78 \\
 VAL & 0.27 & 135.23 & 1.01 & 0.16 & 0.56 & 1.02 & 0.91 & 0.31 & 0.65 & 1.39 \\
 LEU & 0.29 & 142.16 & 0.86 & 0.19 &   -- &   -- &   -- &   -- & 0.64 & 1.28 \\
 ILE & 0.31 & 126.54 &   -- &   -- &   -- &   -- &   -- &   -- & 0.61 & 2.80 \\
 GLY & 0.31 & 116.24 & 0.47 & 2.57 & 0.53 & 0.29 &   -- &   -- & 0.50 & 2.08 \\
 ALA & 0.27 & 110.98 &   -- &   -- & 0.54 & 1.06 & 0.99 & 0.25 & 0.54 & 2.91 \\
 PRO & 0.31 & 121.15 & 0.88 & 0.70 & 0.63 & 0.66 & 1.00 & 0.64 & 0.57 & 2.72 \\
 HIE & 0.27 & 194.74 &   -- &   -- &   -- &   -- &   -- &   -- & 0.56 & 0.52 \\
 HID & 0.23 & 202.98 & 0.51 & 0.74 & 0.55 & 1.56 &   -- &   -- & 0.56 & 3.19 \\
 HIP & 0.27 & 181.40 & 0.50 & 4.47 & 0.58 & 2.28 &   -- &   -- & 0.89 & 1.34 \\
 LYS & 0.27 & 169.68 &   -- &   -- &   -- &   -- & 0.74 & 1.78 & 0.56 & 0.74 \\
 ARG & 0.26 & 126.46 &   -- &   -- & 0.64 & 2.70 &   -- &   -- & 0.51 & 4.14 \\
\hline
\end{tabular}
\caption{Values of the binding energy $\epsilon$ [kJ mol$^{-1}$] 
and the bond length $\sigma$ [nm] for the four ZnO surfaces.
$\sigma$ is measured between the center of mass of an AA and the
surface. The symbol -- signifies non-binding situations.
The superscript $v$ denotes results obtained in vacuum.}
\end{center}
\end{table}

\twocolumn

\clearpage

\underline{{\bf Potential of the mean force at the $(10\bar{1}0)$ ZnO interface in vacuum}}
\vspace*{0.5cm}

\begin{figure}[h!]
\begin{center}
\includegraphics[scale=0.50]{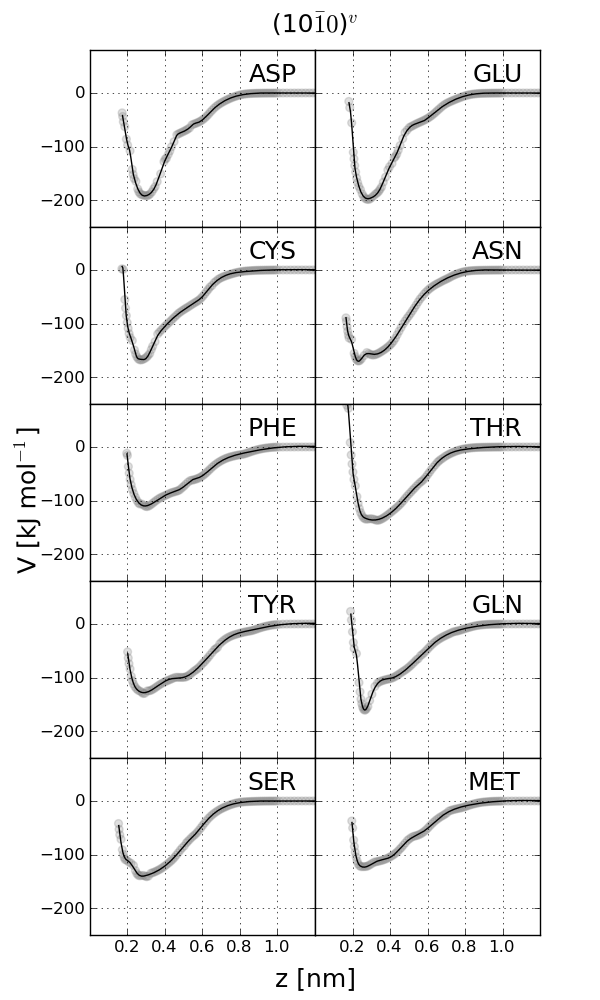}
\caption{$V(z)$ for the AAs in vacuum.}
\label{aa_a_1010}
\end{center}
\end{figure}

\begin{figure}[h!]
\begin{center}
\includegraphics[scale=0.50]{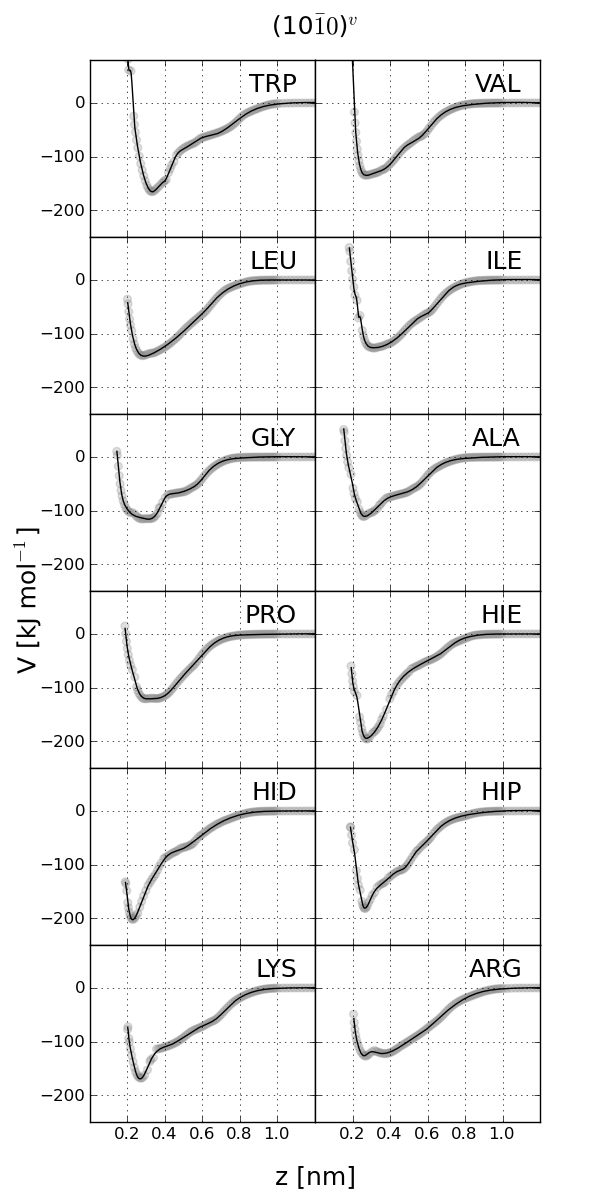}
\caption{$V(z)$ for the AAs in vacuum -- continued.}
\label{aa_b_1010}
\end{center}
\end{figure}

\begin{figure}[h!]
\begin{center}
\includegraphics[scale=0.50]{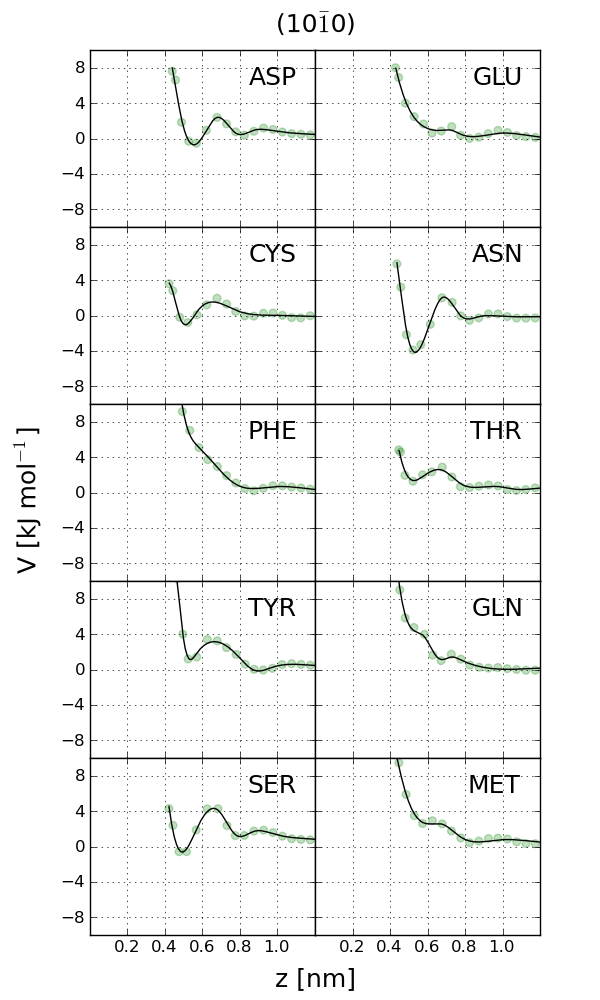}
\caption{$V(z)$ for AAs above the $(10\bar{1}0)$ surface in water.} 
\label{aa_wat_a_1010}
\end{center}
\end{figure}

\begin{figure}[h!]
\begin{center}
\includegraphics[scale=0.50]{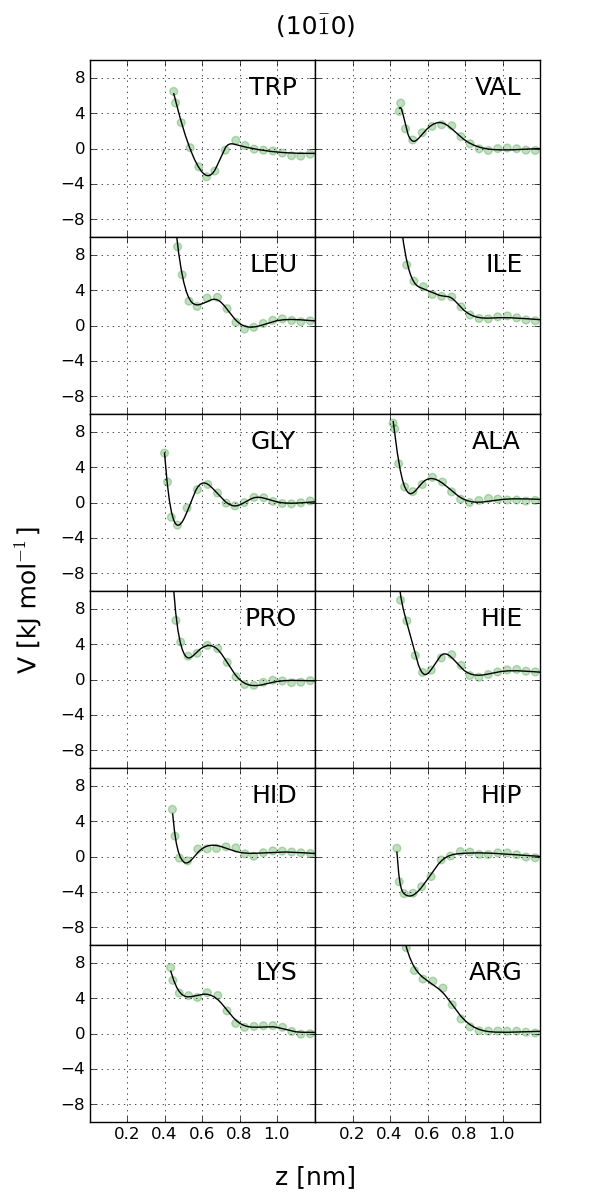}
\caption{$V(z)$ for AAs above the $(10\bar{1}0)$ surface in water -- continued.} 
\label{aa_wat_b_1010}
\end{center}
\end{figure}

\begin{figure}[h!]
\begin{center}
\includegraphics[scale=0.50]{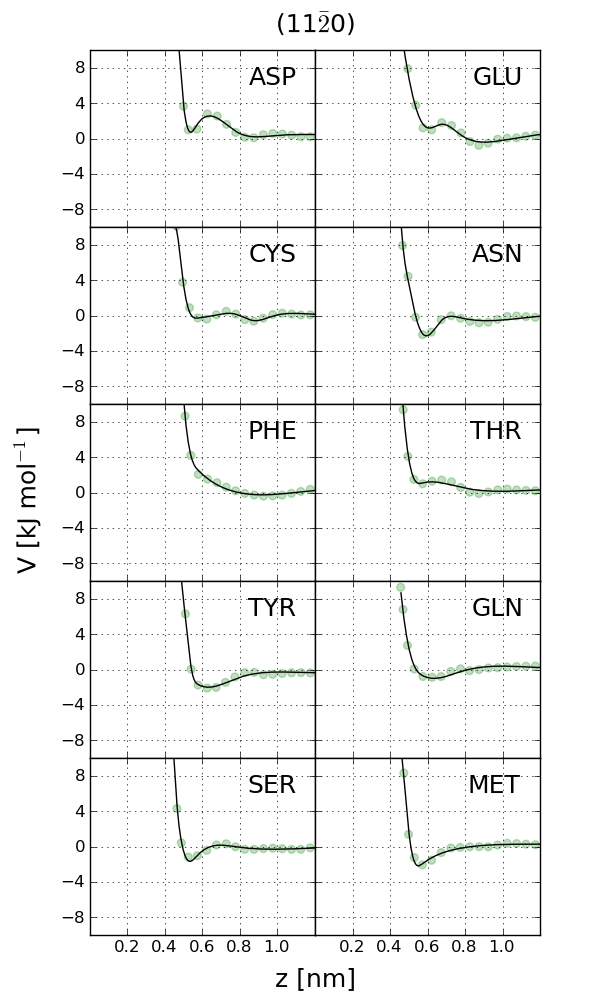}
\caption{$V(z)$ for AAs above the (11$\bar{2}$0) surface in water.} 
\label{aa_wat_a_1120}
\end{center}
\end{figure}

\begin{figure}[h!]
\begin{center}
\includegraphics[scale=0.50]{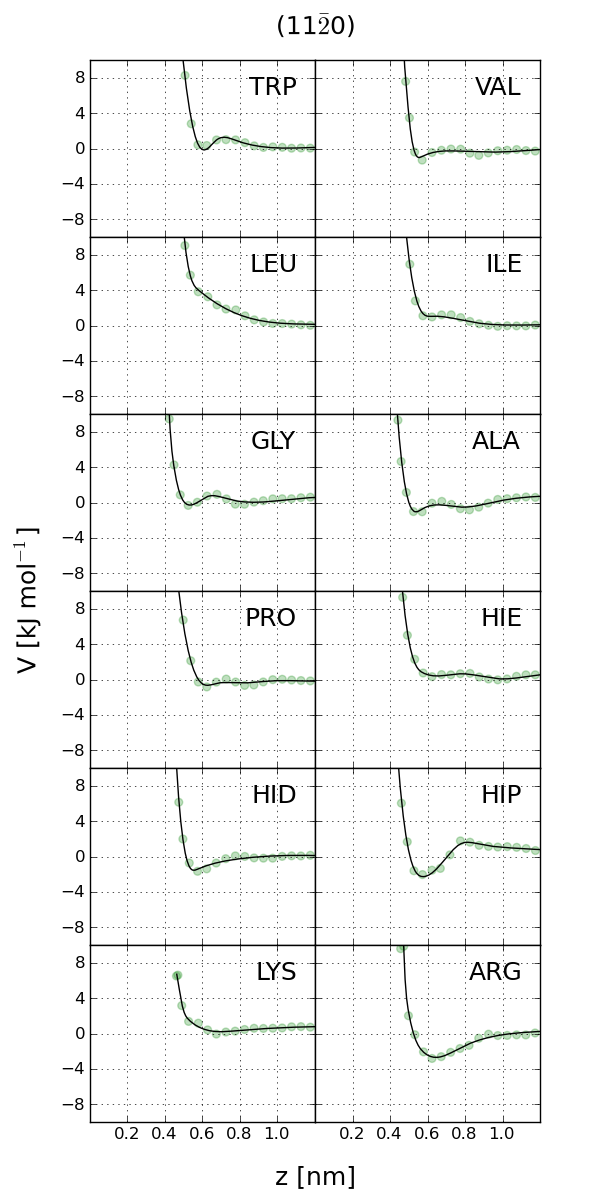}
\caption{$V(z)$ for AAs above the (11$\bar{2}$0) surface in water -- continued.} 
\label{aa_wat_b_1120}
\end{center}
\end{figure}

\begin{figure}[h!]
\begin{center}
\includegraphics[scale=0.50]{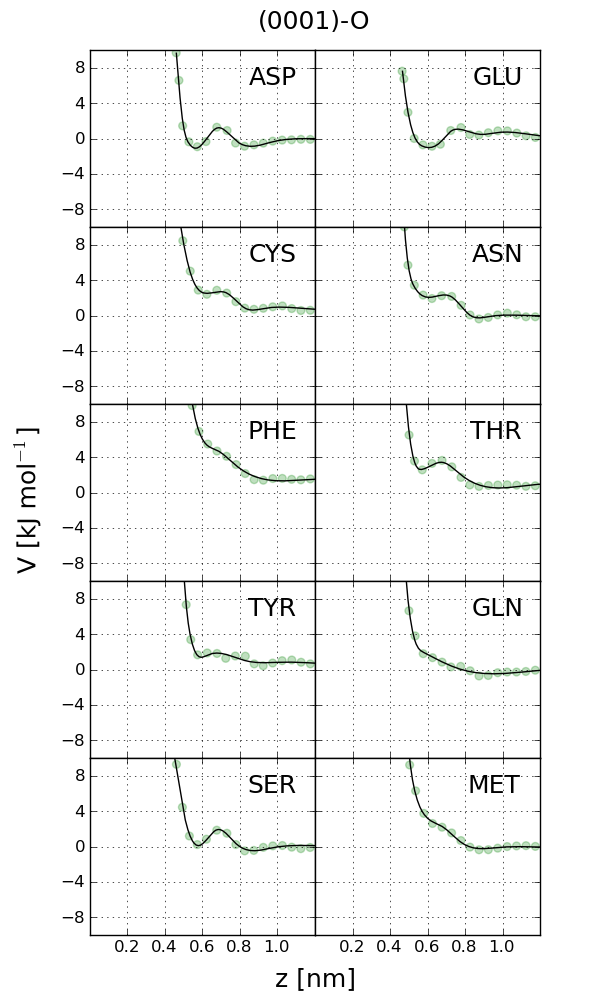}
\caption{$V(z)$ for AAs above the (0001-O) surface in water.} 
\label{aa_wat_a_0001-o}
\end{center}
\end{figure}

\begin{figure}[h!]
\begin{center}
\includegraphics[scale=0.50]{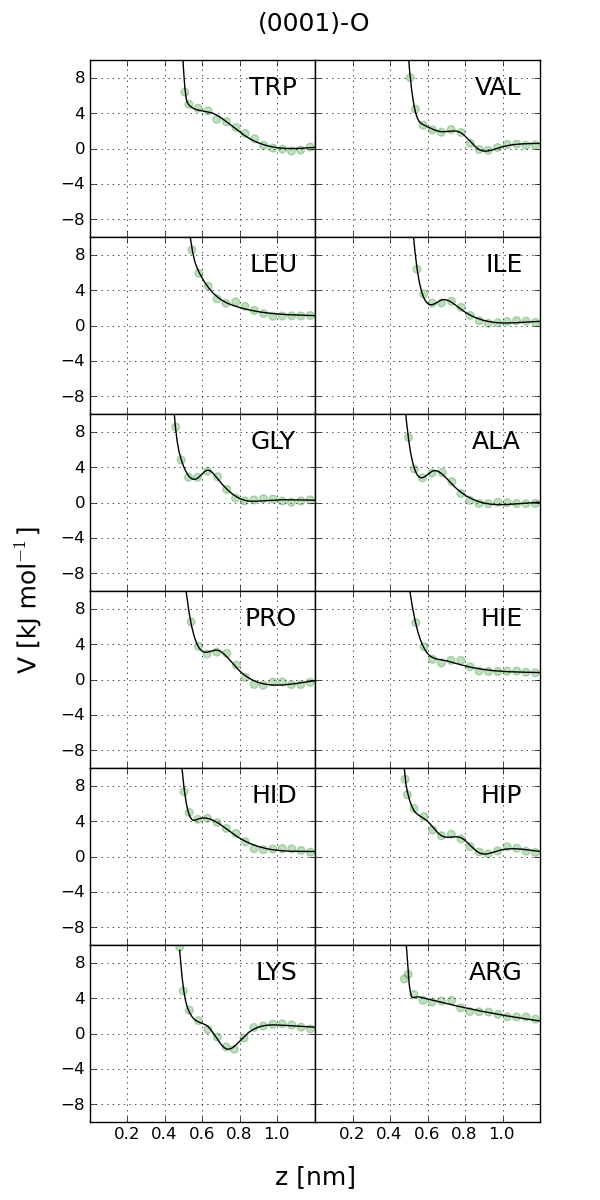}
\caption{$V(z)$ for AAs above the (0001)-O surface in water -- continued.} 
\label{aa_wat_b_0001-o}
\end{center}
\end{figure}

\begin{figure}[h!]
\begin{center}
\includegraphics[scale=0.50]{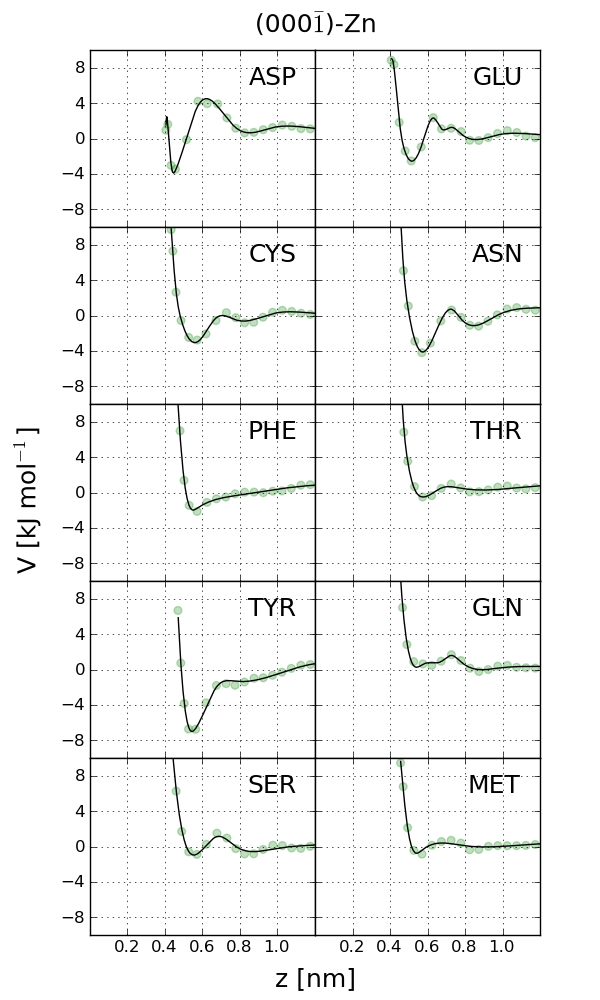}
\caption{$V(z)$ for AAs above the (000$\bar{1}$)-Zn surface in water.} 
\label{aa_wat_a_0001-zn}
\end{center}
\end{figure}

\begin{figure}[h!]
\begin{center}
\includegraphics[scale=0.50]{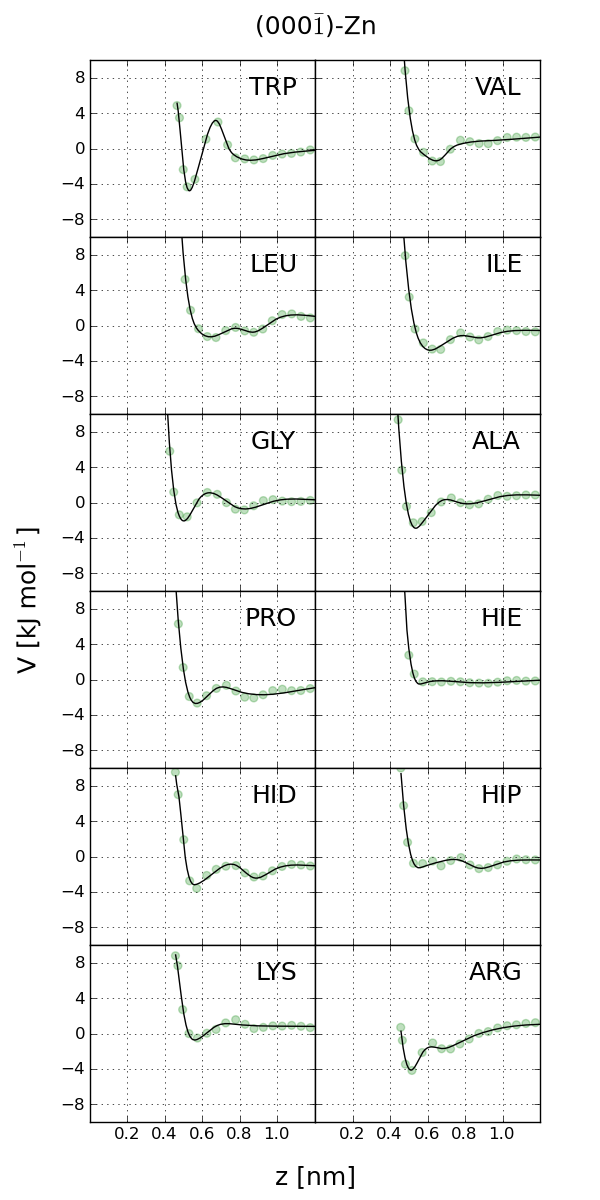}
\caption{$V(z)$ for AAs above the (000$\bar{1}$)-Zn surface in water -- continued.} 
\label{aa_wat_b_0001-zn}
\end{center}
\end{figure}

Here, we provide plots of the PMF obtained by the umbrella
sampling simulations for AAs above the $(10\bar{1}0)$ in vacuum
-- Figures \ref{aa_a_1010} through \ref{aa_wat_b_0001-zn}.
The binding energy varies across the AAs between 100 and 200 kJ/mol 
(Table I in the main text and Table 1 here).
The average bond length is about 0.28 nm.

\clearpage

\underline{{\bf Potential of the mean force at the ZnO interfaces in water}}
\vspace*{0.5cm}
Here, we provide plots of the PMF obtained by the umbrella
sampling simulations for AAs in water
-- Figures \ref{aa_wat_a_1010}, and \ref{aa_wat_b_1010}.
The binding energies are listed in  Table I in the main text and 
Table 1 here.

\clearpage

\newpage

\underline{{\bf Protein 1L2Y at the ZnO surfaces}}
\vspace*{0.5cm}

The following double sets of triple panels provide an analysis of the behavior of 1L2Y during
three examples of adsorption events in analogy to Figures 5 and 6 
in the main text.
We consider events in which at least one atom of the protein is closer than 0.5 nm to
the surface for at least 1 ns.

\clearpage

\begin{figure}[ht!]
\begin{center}
\includegraphics[scale=0.3]{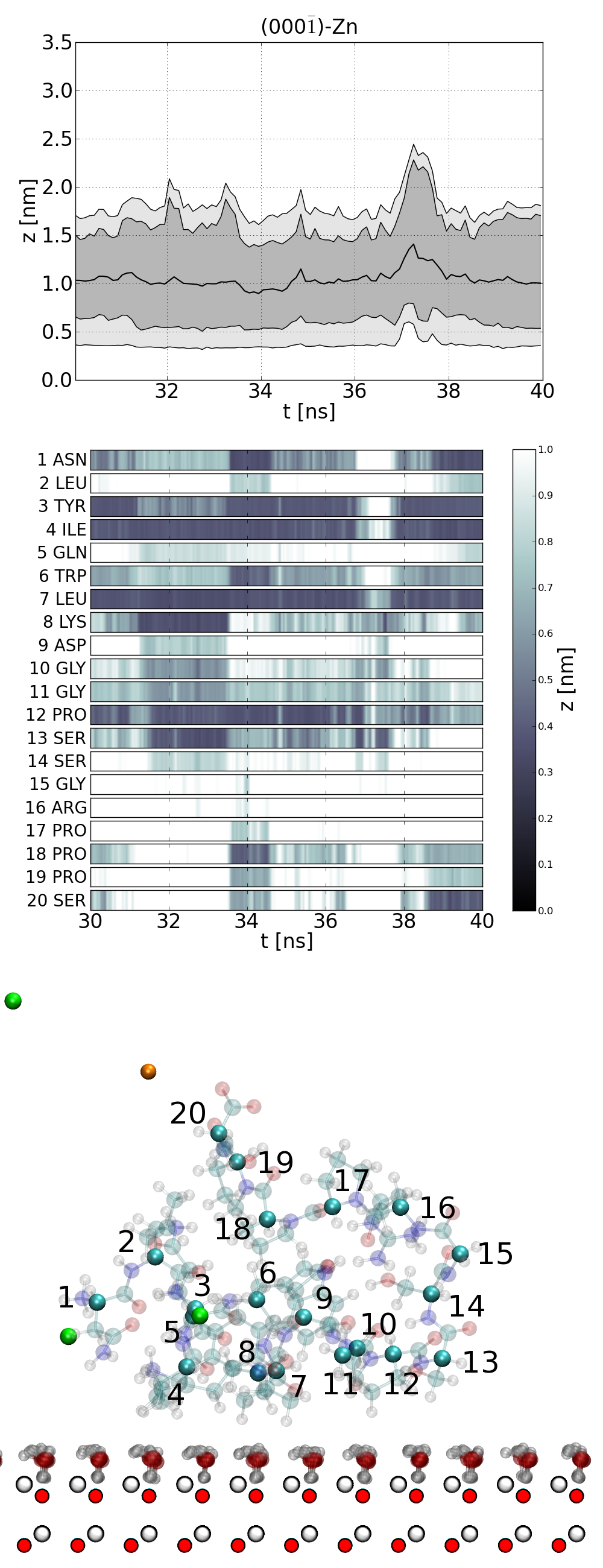}
\caption{The behavior of the tryptophane cage near the (000$\bar{1}$)-Zn surface of ZnO. 
The figure is an analogue of Figure 5  
in the main text. In the selected time interval we identify 3 adsorption events:
31 626 - 33 376 ps, 34 876 - 36 376 ps and 38 876 - 39 876 ps.
The bottom panel shown a snapshot of 1L2Y within the first event (32 000 ps).}
\label{try_0001-zn_a}
\end{center}
\end{figure}

\begin{figure}[ht!]
\begin{center}
\includegraphics[scale=0.3]{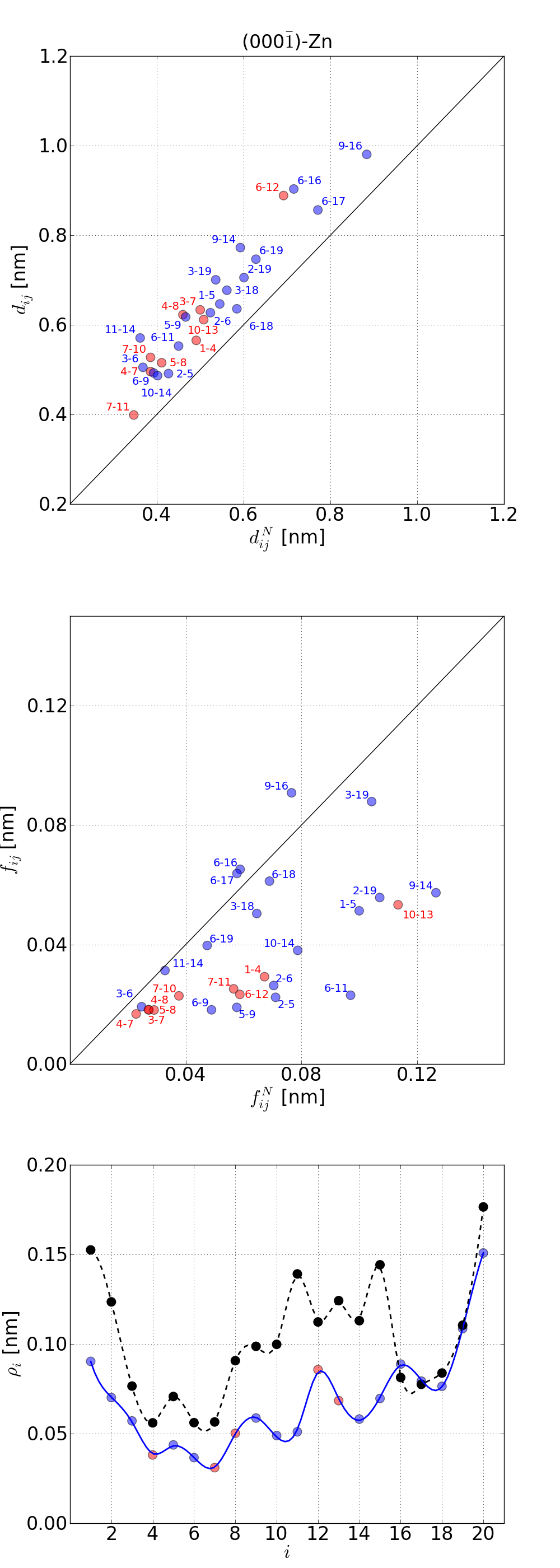}
\caption{
Fluctuational dynamics of a 1L2Y anchored to the (000$\bar{1}$)-Zn surface.
The lines and symbols are as in Figure 6 
in the main paper. 
The first binding event of Figure \ref{try_0001-zn_a} is analyzed.
It involves five AAs (4, 7, 8, 12, and 13).}
\label{try_0001-zn_b}
\end{center}
\end{figure}
\begin{figure}[ht!]
\begin{center}
\includegraphics[scale=0.3]{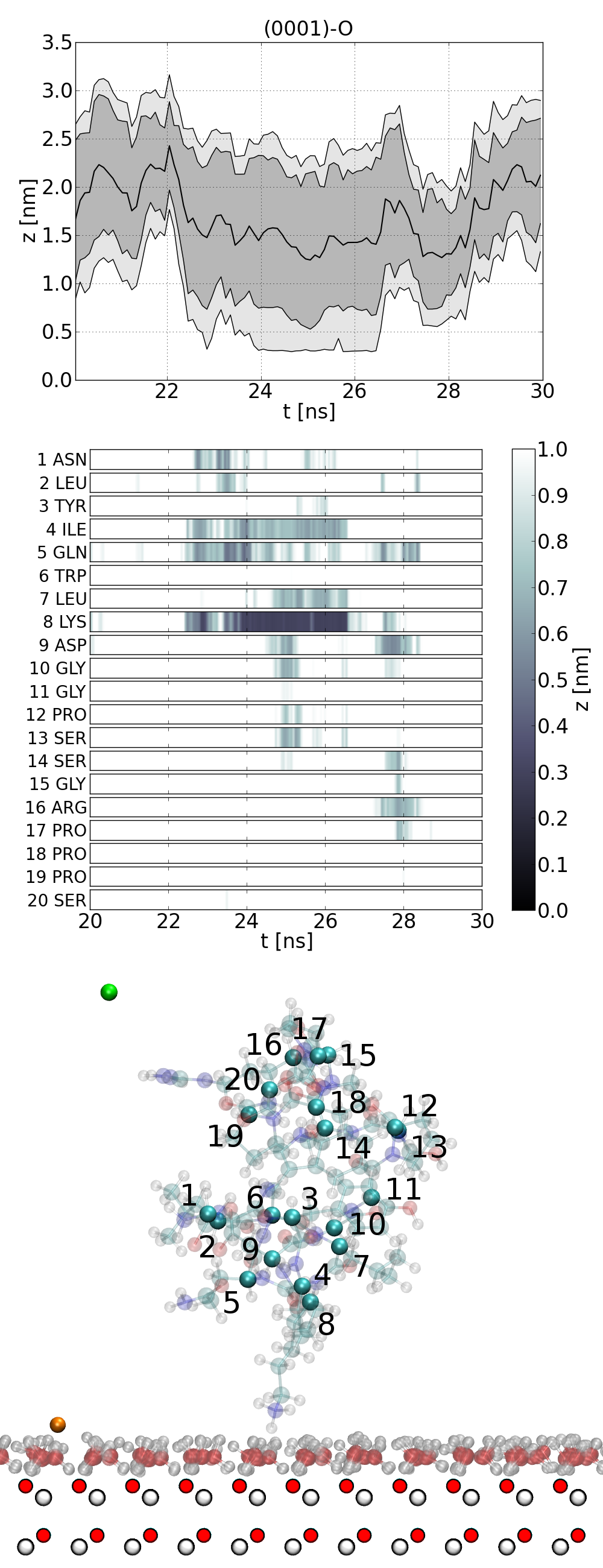}
\caption{Similar to Figure \ref{try_0001-zn_a} but for the (0001)-O surface.
In the selected time interval we recognize adsorption event between:
23 876 and 26 376  ps.
The bottom panel shows a snapshot of 1L2Y at 24 250 ps.}
\label{try_1010_a}
\end{center}
\end{figure}

\begin{figure}[ht!]
\begin{center}
\includegraphics[scale=0.3]{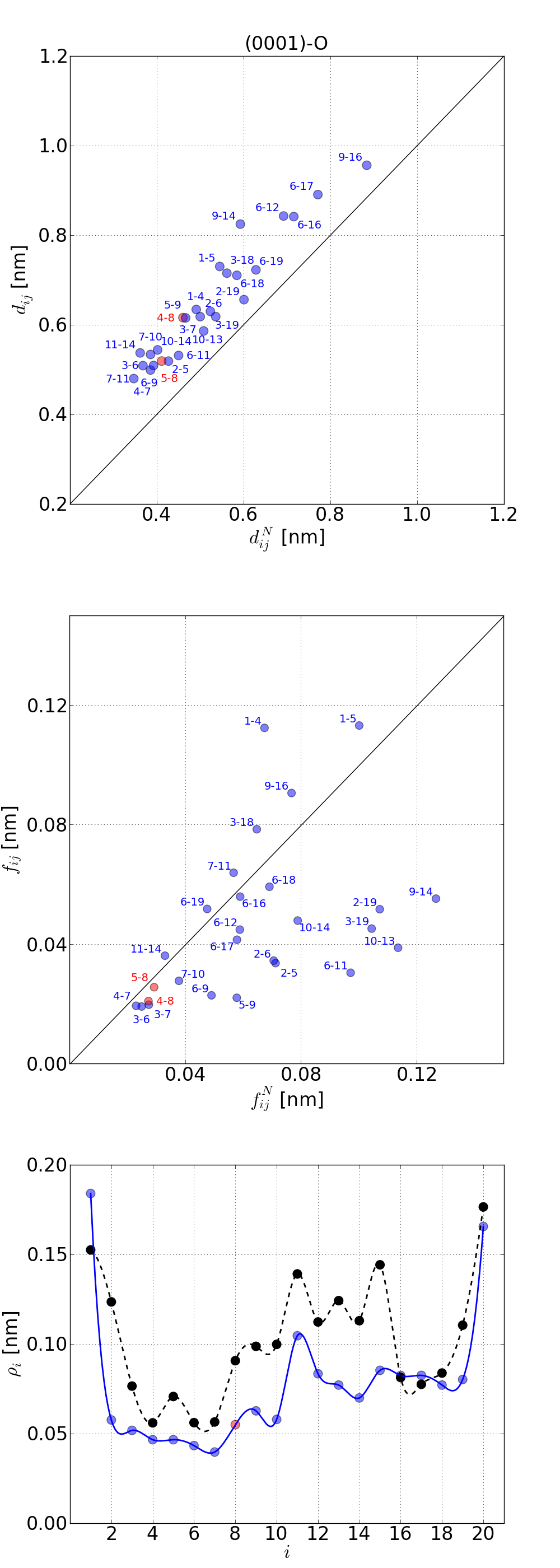}
\caption{Similar to Figure \ref{try_0001-zn_b}. 
The averages are over the time interval between 23 876 and 26 376 ps of the event shown in Figure \ref{try_1010_a}.
It involves one AA (8)} 
\label{try_1010_b}
\end{center}
\end{figure}

\begin{figure}[ht!]
\begin{center}
\includegraphics[scale=0.3]{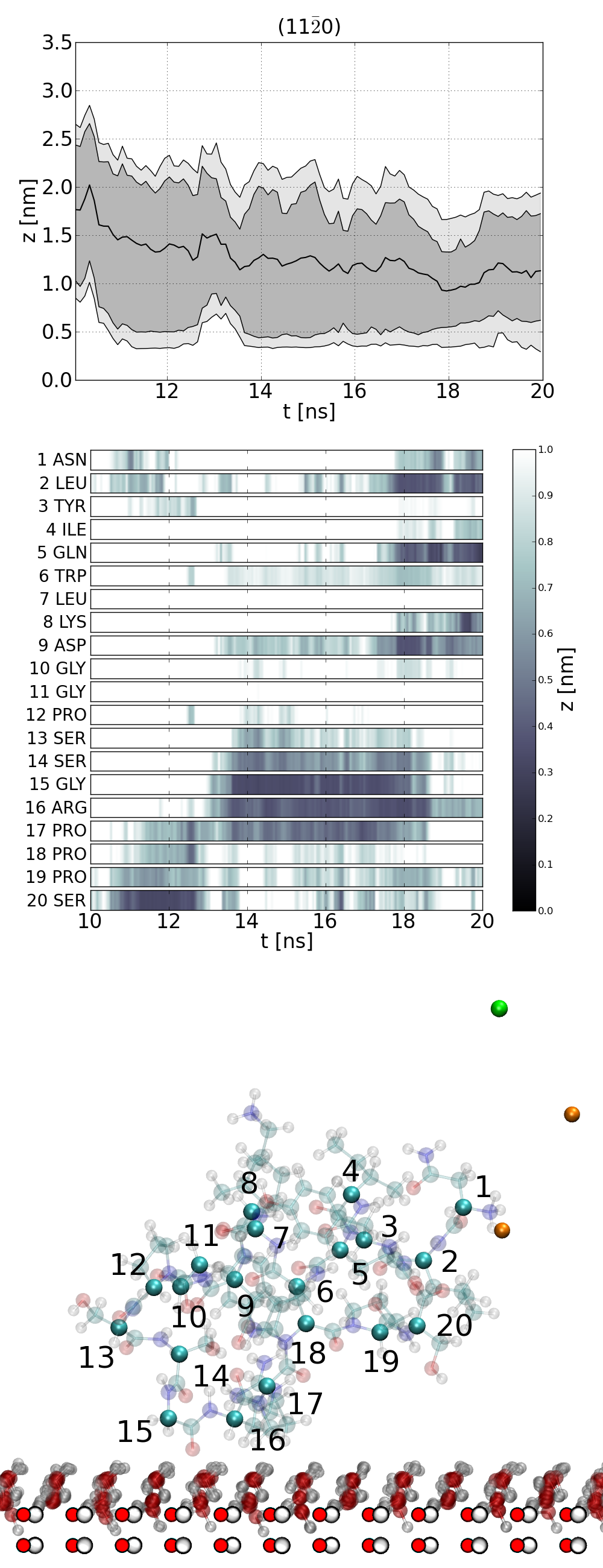}
\caption{Similar to Figure \ref{try_0001-zn_a} but for the (11$\bar{2}$0) surface. 
Between 10 876 and 12 626 only 20-SER is adsorbed. 
Between 15 126 and 17 126 ps the binding event involves three AAs (15, 16 and 17).
The bottom panel shows a snapshot of 1L2Y at 16 000 ps.
}
\label{try_1120_a}
\end{center}
\end{figure}

\begin{figure}[ht!]
\begin{center}
\includegraphics[scale=0.3]{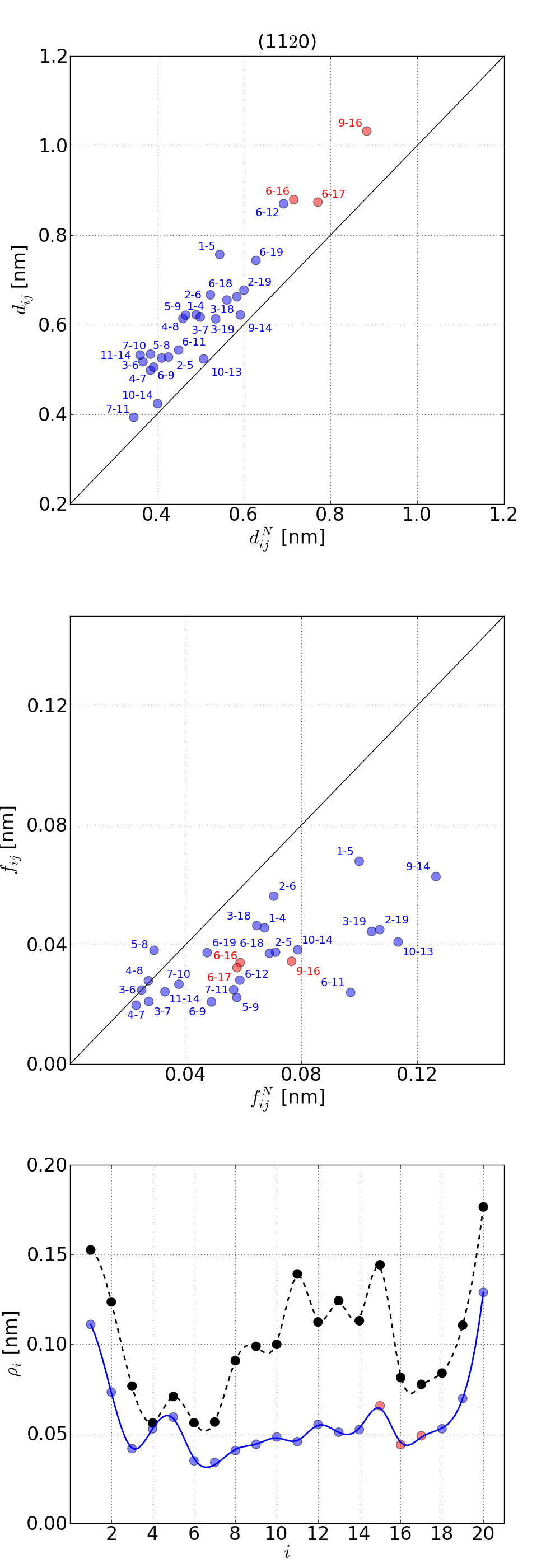}
\caption{ Similar to Figure \ref{try_0001-zn_b}. 
The averages are over the time interval between 15 126 and 17 126 ps 
corresponding to the three-AA event in Figure \ref{try_1120_a}.
} 
\label{try_1120_b}
\end{center}
\end{figure}

\end{document}